\documentclass[sigconf]{acmart}

\usepackage{tabularx}
\usepackage{booktabs} 
\usepackage{geometry}
\usepackage{threeparttable}
\usepackage{array}
\usepackage{titlesec}
\usepackage{CJKutf8}
\AtBeginDocument{%
  }

\copyrightyear{2026}
\acmYear{2026}
\setcopyright{cc}
\setcctype{by-nc-nd}
\acmConference[ASSETS '26]{The 28th International ACM SIGACCESS Conference on Computers and Accessibility}{October 25--28, 2026}{Vila Nova de Gaia, Portugal}
\acmBooktitle{The 28th International ACM SIGACCESS Conference on Computers and Accessibility (ASSETS '26), October 25--28, 2026, Vila Nova de Gaia, Portugal}
\acmDOI{10.1145/3797867.3828983}
\acmISBN{979-8-4007-2521-0/2026/10}

\begin{document}

\title[Supporting Embodied String Learning for Musicians with Blindness and Low-Vision]{Designing for What Cannot Be Seen: Supporting Embodied String Learning for Musicians with Blindness and Low-Vision}

\author{Shi Shi}
\authornotemark[1]
\affiliation{
  \institution{Jacobs School of Music, Indiana University Bloomington}
  \city{Bloomington}
  \state{IN}
  \country{USA}
}
\email{shi14@iu.edu}

\author{Lingyun Chen}
\authornote{Both authors contributed equally to this research.}
\affiliation{
  \institution{Luddy School of Informatics, Computing, and Engineering, Indiana University Bloomington}
  \city{Bloomington}
  \state{IN}
  \country{USA}
}
\email{lch2@iu.edu}

\author{Zitao Zhang}
\affiliation{
  \institution{Luddy School of Informatics, Computing, and Engineering, Indiana University Bloomington}
  \city{Bloomington}
  \state{IN}
  \country{USA}
}
  \email{zhangzit@iu.edu}
  
\author{Amanda R. Draper}
\affiliation{
  \institution{Jacobs School of Music, Indiana University Bloomington}
  \city{Bloomington}
  \state{IN}
  \country{USA}
}
\email{ardraper@iu.edu}

\author{Eli Blevis}
\affiliation{
  \institution{Luddy School of Informatics, Computing, and Engineering, Indiana University Bloomington}
  \city{Bloomington}
  \state{IN}
  \country{USA}
}
  \email{eblevis@iu.edu}
  
\renewcommand{\shortauthors}{Shi, Chen et al.}

\begin{abstract}
Bowed string instruments demand fine-grained bodily coordination that is typically taught through visual demonstration, creating persistent barriers for musicians with blindness and low-vision (BLV). To understand these challenges and explore new design opportunities, we conducted a design study with four advanced string musicians with BLV and three of their instructors. Our team, spanning violin performance and music education, disability studies in music, HCI design, and engineering employed a qualitative, multi-method approach including practice video analysis, lesson observation, expert interviews. Our analysis identifies recurring difficulties in right-hand bow control, left-hand coordination, score access, and memory-intensive practice. Building on these findings, we conducted an exploratory design ideation phase informed by empirical findings and feedback from one musician with BLV. We developed speculative design directions that could potentially address identified breakdowns, while acknowledging that these concepts require further evaluation with instructors and in deployed contexts.
\end{abstract}



\maketitle
\section{Introduction}
Bowed string instruments demand continuous, fine-grained bodily calibration \cite{allingham2021motor, rosenkranz2009regaining}. Players need to synchronize two hands with asymmetric roles: the right hand continuously regulates bow hold, bow path, contact point, pressure, and speed \cite{schoonderwaldt2009mechanics}, while the left hand coordinates finger placement, wrist–forearm alignment, arm rotation across strings, shifting, and vibrato \cite{schoonderwaldt2011mastering, baader1999bimanual, wiesendanger2012fingering}. In typical string pedagogy, these skills are taught through visually grounded practices such as demonstration and mirror-based self-monitoring \cite{galamian2013principles}—players are routinely asked to check whether the bow travels parallel to the bridge or whether the wrist collapses during shifting \cite{fischer1997basics, schoonderwaldt2009mechanics}.

For musicians with blindness and low-vision (BLV), however, these moment-to-moment visual reference points are limited or unavailable. As a result, technique learning is reconfigured from seeing to inferring: players rely more heavily on proprioception, auditory feedback, kinesthetic memory, and teacher mediation to infer where the bow is traveling, how the hand is shaped, and whether posture remains stable \cite{baker2017insights}. Studying string learning with BLV therefore makes visible the implicit visual assumptions of conventional pedagogy and highlights a broader accessibility challenge: in embodied domains, accessibility is often not only about accessing instructions or representations, but about accessing ongoing perceptual reference during action.

Throughout this work, we use language such as "musicians with blind and low-vision" to center disability as a constitutive aspect of musicians' identities and expertise \cite{sharif2022should}, reflecting our commitment to designing with disabled people as agents rather than recipients of support. Drawing on critical disability studies in HCI \cite{spiel2020nothing, mankoff2010disability, yin2026active}, we treat musicians with BLV as knowledgeable practitioners whose lived experience helps surface assumptions embedded in visually grounded pedagogy.
Prior research in accessible music has established important foundations for notation access, such as tactile or audio-mediated scores \cite{zhang2025accessibility, lu2022learning, lu2023there, langolff2000mfb}, accessible interfaces for music production and rehearsal \cite{payne2020blind}, and broader accounts of how musicians with BLV navigate learning and performance \cite{pedrini2020evaluating, lu2024we}. These works have shown that  music learning for people with BLV is not simply a matter of reading scores, but a broader practice ecology shaped by notation, memory, touch, auditory strategies, and teacher support. At the same time, less attention has been paid to the instrument-specific demands of technique learning in bowed strings, where sound production depends on continuous bodily adjustment and where many important technique states are conventionally taught and corrected through vision. This gap is especially salient in violin and viola learning. Unlike cello or bass, which provide additional physical grounding through instrument size, floor contact, or seated posture, violin and viola are supported primarily by the body, reducing stable external reference points and making spatial calibration highly dependent on visual self-reference.

To address this gap, we ask two questions: (1) What recurring breakdowns do violin and viola musicians with blindness and low vision encounter during technique learning and self-practice? and (2) How can these breakdowns inform design directions for supporting non-visual, practice-compatible calibration in upper-string learning? To answer these questions, we conducted a qualitative multi-method study \cite{palakshappa2006using, lazar2017research} with four string musicians (violin/viola) with BLV and three string instructors directly teaching them. We triangulated across practice video analysis, lesson observation, expert interviews, and audio-based practice diaries to capture learning as an embodied, situated, and temporally distributed process, with particular attention to moments where learners struggled to sense, interpret, or verify technique states during independent practice. 

Across cases, we identified recurring difficulties that extended beyond notation access and were deeply intertwined with technique development and practice routines. These included breakdowns in right-hand bow control (e.g., sensing bow drift, maintaining a relaxed bow hold under fatigue), left-hand coordination (e.g., calibrating wrist–forearm alignment and arm rotation in relation to intonation across string crossings), and the multi-step, memory-intensive workflow of learning, monitoring, and rehearsing repertoire without stable external reference points. Building on these findings, we then conducted a bounded design synthesis to explore how three early concept directions—\textit{Augmented Bow Track}, \textit{augmented bow holder aid}, and \textit{a left-hand alignment wearable tethered} to the instrument—might externalize otherwise invisible technique states into non-visual, practice-compatible cues during self-practice.

This paper makes the following contributions:
\begin{itemize}
    \item An empirically grounded, string-specific account of recurring breakdowns in upper-string learning for people with BLV, focusing on right-hand control, left-hand alignment and intonation, and score-related practice.
    \item Accessibility insights into embodied learning, showing that the central challenge in this domain is not only access to musical information, but access to perceptual reference during action.
    \item Design implications and early concept directions for future accessible music technologies, showing how non-visual, practice-compatible feedback might support embodied learning beyond notation access alone.
\end{itemize}

\section{Related Work}
Our work builds on prior research in accessible music learning and technologies for string practice, reviewed in three parts: BLV music learning as access and adaptation, embodied technique learning beyond access, and existing tools for string musicians with BLV. Existing scholarship has made important progress in documenting how musicians with BLV access musical materials, adapt learning strategies, and participate in pedagogical settings that are predominantly designed around vision. In parallel, a range of pedagogical tools and assistive aids have been developed to support specific aspects of string learning, including bowing, fingering, and score study. Across accessibility research, however, designing with rather than for experts with disabilities has become a foundational methodological and ethical commitment \cite{mankoff2010disability, spiel2020nothing, hofmann2020living}, and this orientation shapes how we approach string learning for people with BLV in this paper, identifying a gap in how non-visual interaction might support continuous, embodied calibration during upper-string practice.

\subsection{Music Learning for people with BLV}
Prior research has established that musicians with BLV face persistent barriers in accessing musical materials and participating in conventional learning environments \cite{payne2025access}, including acquiring accessible scores \cite{lu2023there, lu2023playing}, navigating notation systems \cite{baker2016perceptions, payne2025access}, receiving visually mediated feedback \cite{zhang2025accessibility}, and obtaining sustained pedagogical support from teachers \cite{pino2019teaching, baker2016perceptions}. To address these challenges, learners with BLV rely on tactile exploration, auditory memory, and teacher-mediated adaptation \cite{lu2023there, chen2006tactile, pino2019teaching}—suggesting that music learning for people with BLV is not simply a matter of score access, but a broader practice ecology shaped by notation, memory, touch, auditory strategies, and teacher support.
For example, Payne and An document how musicians with low-vision customize Large Print and Modified Stave Notation by adjusting spacing, contrast, and visual density to make scores usable in practice \cite{payne2025access}, while complementary research emphasizes the role of touch and vibrotactile cues in supporting instrument exploration, instruction following, and routine building \cite{lu2023playing}. Collectively, this literature demonstrates that accessibility in music learning extends beyond information delivery to the ongoing work of translating visually taught cues into usable forms during practice.

At the same time, most of this work has focused on access to musical content, notation workflows, and general learning strategies, rather than the instrument-specific demands of technique learning. This leaves open an important question: how learners with BLV manage forms of musical practice that depend not only on accessing information, but on continuously sensing, monitoring, and adjusting the body during performance.

\subsection{Embodied Technique Learning Beyond Access}
String playing is an embodied skill in which sound production depends on continuous, fine-grained bodily calibration of bow direction, bow--string contact, finger placement, and posture, rather than on discrete lookup of correct positions \cite{galamian2013principles}. In classical violin and viola pedagogy, teachers often diagnose these ongoing technique states through vision: they watch bow trajectory, wrist shape, and left-arm rotation, and use mirrors or video to help students compare, monitor, and correct their own movements \cite{galamian2013principles, schoonderwaldt2009violin}. Unlike instruments anchored to the floor or body, violin and viola offer few stable external reference points, making spatial calibration especially dependent on these visually grounded routines \cite{galamian2013principles, fischer1997basics}. For learners with BLV, this creates a challenge that extends beyond access to notation or instructional materials: it concerns how technique-relevant body states are perceived, monitored, and corrected during action \cite{lu2023there, baker2016perceptions}.

Prior work in motor learning, HCI, and interactive training systems suggests that non-visual feedback can help externalize otherwise invisible performance states during continuous skill acquisition \cite{sigrist2013augmented, effenberg2005movement}. Sonification and haptic or vibrotactile feedback have been used to support trajectory, posture, force, and movement awareness in domains such as rehabilitation, sports, and physical skill learning \cite{effenberg2005movement, schaffert2011acoustic, lieberman2007tikl, spelmezan2012tactile, schonauer2012multimodal}. Within music, several systems have similarly explored real-time technique feedback, including MusicJacket for straight bowing, augmented-reality displays for bow trajectory and intonation, and sensor-based approaches for estimating bow orientation and contact parameters \cite{vanderlinden2011musicjacket, johnson2010musicjacket, pardue2019realtime, dalmazzo2021realtime, larkin2008sonification, giordano2021assessing}.
However, these systems have been developed largely for sighted learners and within practice models that still assume access to visual confirmation. They often combine haptic or auditory feedback with screens, visual displays, or teacher-led interpretation, and are rarely evaluated with musicians who are BLV or in the self-directed, hands-busy practice contexts that characterize much of learning \cite{lu2023there, lu2022learning}. As a result, prior work on music learning for people who are BLV has largely centered access, adaptation, and pedagogical support, while prior work on embodied technique augmentation has centered sighted motor learners—our work sits at this intersection, asking how technique states typically monitored through vision might instead be perceived and adjusted during practice without visual self-reference.

\subsection{Tools for String Musicians with BLV}
A variety of tools and pedagogical aids have been developed to support string players with BLV across right-hand bowing, left-hand fingering and alignment, and score study. Together, these supports compensate for the reduced availability of visual feedback through mechanisms such as physical stabilization, tactile marking, auditory reference, and step-by-step access. While valuable, many are most effective for early-stage learning or targeted verification rather than for supporting ongoing, in-action calibration during hands-busy practice.

\begin{figure}[h]
    \centering
    \includegraphics[width=0.75\linewidth]{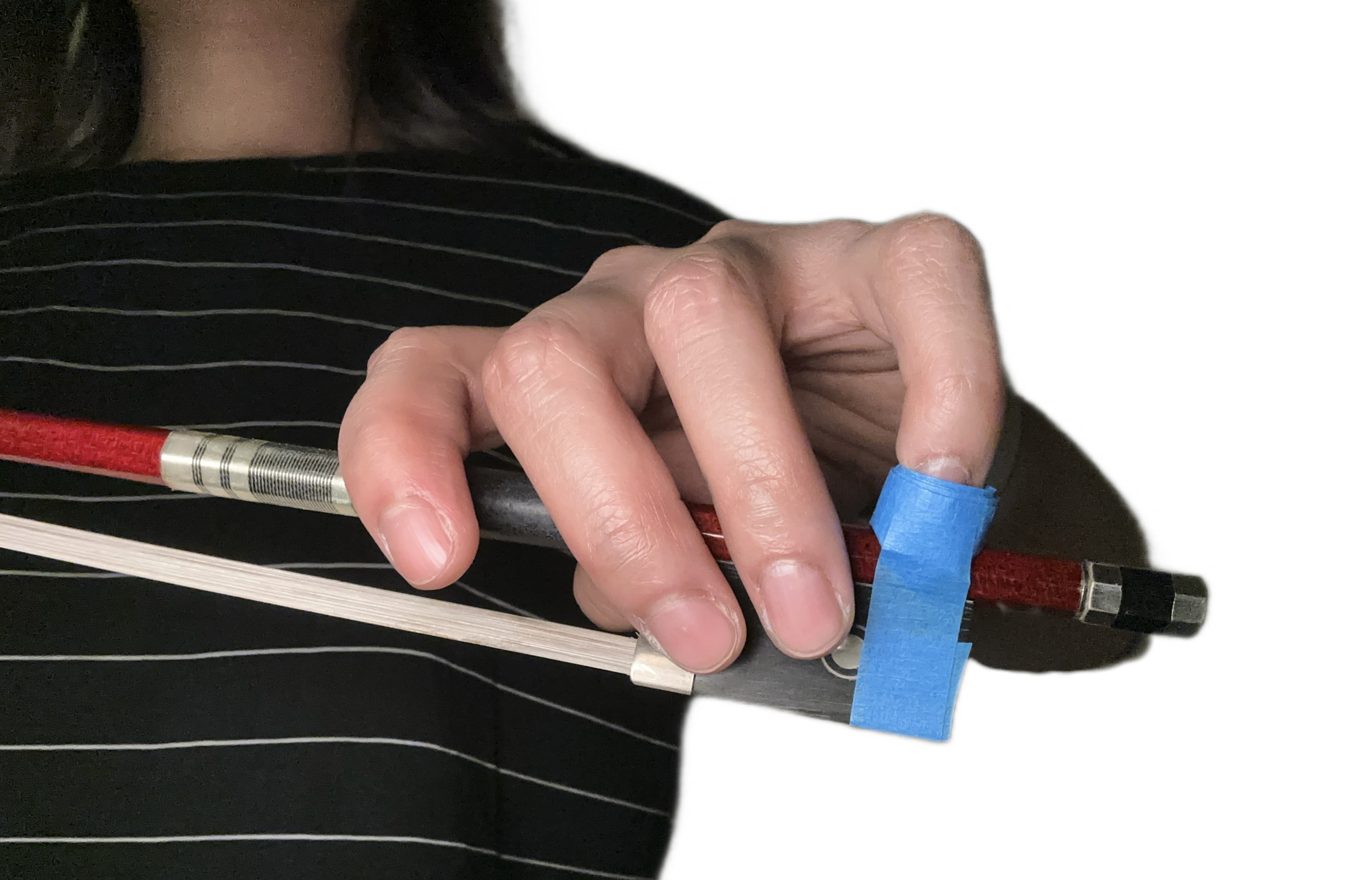}
    \caption{Pinky house using masking tape}
    \label{fig:BLV RH tools}
    \Description{Two pedagogical aids for the right-hand bow hold used in beginner string teaching. On the left, a commercial pinky hold: a small rubber cap that clips onto the bow stick near the frog with a curved resting slot for the pinky. On the right, a DIY pinky house: rubber bands and a folded-paper sleeve tied around the same region of the bow to create an improvised finger rest. Both devices physically constrain where the right-hand pinky contacts the bow so the finger does not collapse or slip during early practice.}
\end{figure}

Many of these tools support isolated components of learning rather than helping learners perceive and interpret technique states as movement unfolds \cite{lu2023there, lu2022learning, goldstein2000music}. For right-hand support, bow-hold assistants such as \textit{pinky holds} (and DIY \textit{``pinky houses''}, Figure~\ref{fig:BLV RH tools}) attach near the frog to support pinky placement and reduce collapse \cite{zweig_stringpedagogy_2025}. These aids are particularly useful in early training; as technique develops, however, players rely on subtle pressure redistribution and balance adjustments for tone, dynamics, and articulation, which fixed supports may restrict. Related \textit{bow grip aids} similarly stabilize the overall hold through fixed ``finger homes,'' but may inhibit the finger-pressure redistribution required in more advanced playing. A different class of tools targets bow trajectory \cite{rolland1979movement}: bow-guiding aids form a narrow track between the bridge and fingerboard to keep the bow in a stable lane, helping beginners build basic habits but preventing error mechanically rather than helping players perceive drift, tension, or balance while playing.

For the left hand, technique is closely tied to intonation \cite{galamian2013principles, reel2004strings, mishra2000questions, rolland1974teaching}. Because pitch accuracy depends on finger placement, hand--arm configuration, and stylistic context \cite{morrison2011intonation, rolland1974teaching}, string pedagogy has developed tools for early fingerboard calibration \cite{shipps2014top, sarch1996violin}. Finger tapes \cite{bergonzi1997effects, huovinen2024string} on the fingerboard provide visual or tactile landmarks that help learners build a basic spatial map of note locations \cite{termeer_fretted_violin_nodate}, while customized fretted violins offer clearer touch-based reference points. These tools can support beginners, but they may also encourage over-reliance on fixed landmarks rather than listening-based adjustment \cite{chen2008pitch, hafke2016violinists}. Electronic tuners are also widely used \cite{zabanal2019effects}; accessible modes can provide sustained reference pitches for ear-based matching, but they do not capture the context-sensitive intonation goals that emerge with experience \cite{mcgarry1984equal, duffin2007equal, whitcomb2017intonation}. Wrist and arm alignment are also critical for technique development and injury prevention \cite{visentin2015unraveling, mccrary2016effects}. Wrist practice aids can provide external stabilization during early training \cite{berend1972interrelation}, but they may also restrict the mobility needed for shifting and vibrato. Overall, left-hand tools support early awareness and placement, but offer limited support for ongoing alignment and intonation calibration during independent practice.

Beyond hand technique, score study involves not only identifying notes and rhythms \cite{ellis1994selected, hunsberger1992art, lane2006undergraduate}, but also interpreting performance instructions such as bowings, fingerings, dynamics, and articulation. Audio recordings are a common entry point because they provide an aural model for tempo, phrasing, and expression, but they become insufficient as repertoire grows denser or rhythmically more complex. Score-reading software, often used alongside screen readers, offers another pathway for accessing written notation: applications such as MuseScore and Sibelius, as well as LilyPond-generated outputs, can provide access to pitch, duration, measure numbers, dynamics, and other markings, while enlargement features support low-vision learners. However, screen-reader-based score study imposes substantial cognitive and interactional overhead: navigation is often slow, linear, and mentally demanding, making these tools well suited to targeted verification but less suited to fluid, real-time score study during hands-busy practice.

Across right-hand, left-hand, and score-study supports, existing tools are valuable but partial: they help learners access information, stabilize early technique, or prevent certain errors, yet provide limited support for perceiving and interpreting ongoing technique states in the flow of everyday practice. This gap motivates our focus on how musicians with BLV manage technique breakdowns in upper-string learning, and on how design might support non-visual, continuous, and practice-compatible forms of feedback.

\section{Methods}
We employed a design-oriented, qualitative, multi-method approach \cite{palakshappa2006using, lazar2017research, adams2008qualititative} to examine how string musicians with BLV learn, practice, and receive instruction. Our methodological orientation treats learning as an embodied, situated, and temporally distributed process, rather than as isolated moments of performance. In Stage 2, we used an exploratory design ideation approach informed by Research through Design (RtD) principles \cite{bowers2012logic, zimmerman2007research} and speculative design practices \cite{dunne2024speculative, chen20253r}. We treated empirical findings as design inspiration and used iterative concept sketching to explore speculative yet grounded design directions for BLV string learning \cite{chen2026robots}.

We combined participant-generated practice artifacts, lesson observations, and in-depth interviews with both students and instructors. These empirical materials informed an interdisciplinary speculative synthesis process, through which recurring technique breakdowns were translated into design concepts aimed at supporting independent practice.

\subsection{Participants}
Our study involved two groups of participants: string musicians with BLV and string instructors with direct experience teaching learners with BLV. This combination enabled us to examine string learning from both first-person practice perspectives and instructional standpoints, situating individual technique challenges within broader pedagogical contexts. Participants were recruited through disability communities and personal networks.

All participants are referred to using pseudonyms in this paper. During the consent process, participants were informed about how their data would be reported, including de-identification and the presentation of attributed quotations. All participants provided informed consent prior to participation, and this reporting approach was reviewed and approved as part of our Institutional Review Board (IRB) protocol.

\subsubsection{String Musicians with BLV.}

We collaborated with four string musicians (violin and viola) with BLV, with diverse musical backgrounds, vision conditions, and levels of training. All participants were actively engaged in regular practice at the time of the study, either as students in formal music programs or as advanced musicians. Their ages ranged between 25 and 34 years. Demographic information, musical backgrounds, and vision conditions are summarized in Table~\ref{tab:p_stu}.

\begin{table*}[t]
\caption{Participant Details, Musicians with BLV}
\label{tab:p_stu}
\centering
\begin{tabularx}{\textwidth}{l c c >{\raggedright\arraybackslash}X c p{3cm}}
\toprule
ID & Age & Gender & Description & Yrs of Exp & Visual Impairment \\
\midrule
M1 & 26 & M & Doctoral student in viola performance at a large public university in the U.S. Midwest & 13 & Fully blind \\
M2 & 28 & M & Violin performance master's student at a large public university in the U.S. Southwest & 10 & Fully blind \\
M3 & 25 & M & Recent graduate in music education from a large public university in the U.S. South & 19 & Low vision \\
M4 & 34 & M & Collegiate-level violin teacher in China & 24 & Fully blind \\
\bottomrule
\end{tabularx}
\end{table*}

\subsubsection{Instructors.}

We involved three string instructors with direct experience teaching learners with BLV. Two instructors were directly associated with participants M1 and M2; one was a teaching assistant affiliated with M1's instructor. These instructors provided perspectives on pedagogical adaptation, instructional challenges, and the demands of teaching string technique without relying on visual demonstration. Demographic information and teaching backgrounds are summarized in Table~\ref{tab:p_ins}.

\begin{table*}[t]
\caption{Participant information: instructors.}
\label{tab:p_ins}
\centering
\small
\renewcommand{\arraystretch}{1.1}
\setlength{\tabcolsep}{4pt}
\begin{tabularx}{\textwidth}{@{}l c c p{1.7cm} X p{2.4cm}@{}}
\toprule
\textbf{ID} & \textbf{Age} & \textbf{Gender} & \textbf{Experience} & \textbf{Occupation} & \textbf{Relationship to BLV participants} \\
\midrule
I1 
& 76 
& M 
& More than 50 years 
& Viola professor at a large public university in the U.S. Midwest; former viola principal of a major symphony orchestra 
& M1’s instructor \\

I2 
& 55 
& F 
& More than 30 years 
& Violin professor at a large public university in the U.S. Southwest; concertmaster of the local symphony orchestra
& M2’s instructor \\

I3 
& 23 
& M 
& One year 
& Teaching assistant in viola performance at a large public university in the U.S. Midwest 
& Teaching assistant to M1’s instructor \\

\bottomrule
\end{tabularx}
\end{table*}

\subsection{Empirical Study of String Learning for People with BLV (Stage 1)}

Our first-stage empirical study examined how string musicians with BLV learn, practice, and receive instruction in everyday contexts. We adopted a multi-stage qualitative process designed to capture string learning as an embodied, situated, and temporally distributed practice, moving from everyday self-directed practice, to pedagogical interaction, and finally to participant and instructor reflection. This progression enabled us to connect observed bodily actions with lived experience and instructional intent, capturing technique breakdowns as they occurred across different learning contexts.

\subsubsection{Practice Video Collection.}

Each of the four musicians provided video recordings of typical practice sessions (10--45 minutes; 90 total minutes). We analyzed these videos to observe embodied techniques—bow trajectory, hand coordination, posture—and self-developed strategies, examining how technique breakdowns manifested in action beyond what participants could articulate in interviews.

\subsubsection{Lesson Observation.}

We conducted one 60-minute lesson observation each with M1 and M2 (120 total minutes). Additionally, we observed one chamber rehearsal with M1, the viola DMA student (120 total minutes).M3 and M4 were not at the time of the study. M3 had recently graduated and was no longer taking regular lessons, and M4 worked as an independent collegiate-level teacher rather than as a student—so no comparable lesson observation was available for either. Both instead contributed practice videos and interview data. During observations, we focused on how teachers communicated technique, delivered feedback when visual demonstration was unavailable, and used physical guidance to convey information. This revealed the interactional burden on teachers and situated learning challenges within broader pedagogical contexts.

All observations were audio-recorded. All researchers kept field notes using a structured template tracking instructional language, feedback timing, physical guidance, and moments of visual assumption \cite{emerson2011writing}, expanded within 24 hours.

\subsubsection{Interviews.}

Semi-structured interviews \cite{roulston2010reflective} were conducted with all participants (four musicians: 90 minutes each; three instructors: 45 minutes each; 495 total minutes). Musician interviews explored learning histories, practice strategies, technique development, and use of assistive technologies. Instructor interviews focused on pedagogical adaptations, instructional challenges, and the 
burden of teaching without visual demonstration. Interviews were scheduled 1--2 weeks after lesson observations to allow reflection and connect lived experience with observed behavior. All interviews were audio-recorded and fully transcribed.

\subsubsection{Data Analysis.}

All materials were analyzed through iterative thematic analysis \cite{terry2017thematic}.Interview transcripts and observation notes were open-coded for technique 
sensing, feedback delivery, and learning barriers. To ensure coding 
consistency, 20\% of transcripts and observation notes were independently 
coded by two researchers, achieving 85\% inter-rater agreement on primary 
codes. Practice videos were analyzed through embodied notation—identifying recurring moments and creating brief summaries (2--3 sentences)—then coded alongside interviews and observations to triangulate what participants said, what was observable, and what instructors identified as problems. Codes were clustered into higher-level themes capturing recurring tensions (e.g., absence of continuous perceptual feedback) that directly informed Stage 2. Emerging interpretations were revisited with participant feedback to ensure grounding in lived experience \cite{patton2014qualitative}.

\subsection{Exploratory Design Ideation Informed by Research through Design Principles (Stage 2)}

Following the empirical study, we conducted an exploratory design ideation phase informed by Research-through-Design (RtD) principles \cite{bowers2012logic, zimmerman2007research}. Rather than conducting full participatory co-design, we adopted a bounded, researcher-led synthesis in which empirical findings were treated as generative material for speculative concept development \cite{chen2026robots}. To ground emerging concepts in lived experience, we invited one musician with BLV (M2) to provide feedback during concept refinement, recognizing that full multi-participant iteration was not feasible given scheduling and health constraints during the study window.

\begin{table}[t]
  \caption{Research team members and roles in design ideation.}
  \label{tab:team-roles}
  \centering
  \small
  \renewcommand{\arraystretch}{1.08}
  \setlength{\tabcolsep}{4pt}
  \begin{tabularx}{\columnwidth}{@{}l p{2.3cm} X@{}}
    \toprule
    \textbf{ID} & \textbf{Background} & \textbf{Role in RtD process} \\
    \midrule
    A1 & Engineering 
       & Led technical exploration and drafted initial design concepts based on empirical findings. \\
    A2 & String pedagogy 
       & Proposed and refined design ideas based on professional experience as a string instructor. \\
    A3 & Design 
       & Developed visual sketches and examined concepts from a UX perspective. \\
    A4 & HCI \& accessibility research 
       & Synthesized design implications and assessed interactional coherence. \\
    \bottomrule
  \end{tabularx}
\end{table}

The design process was led by a cross-disciplinary author team with expertise in string pedagogy, disability music education, interaction design, sensing technologies, engineering, and HCI. We used Stage 1 findings as generative material for design exploration, focusing on recurring challenges in sensing technique states, coordinating complex embodied actions, and sustaining practice without continuous visual reference or instructor oversight. Early ideation sessions were intentionally exploratory, and were not constrained by immediate technical feasibility, as the goal of this phase was to probe the design space rather than develop deployable systems.

To ground this process in lived experience, we invited one musician with BLV from Stage 1 (M2) to participate in follow-up feedback sessions during concept refinement. M2 was selected because he was actively engaged in regular practice, had reflected in detail on technique breakdowns during the empirical study, and was available for multiple rounds of discussion. His role in Stage 2 was not to co-lead concept generation, but to provide situated feedback on embodied feasibility, pedagogical fit, likely use in practice, and possible risks. Because this phase was intended to refine early design directions through iterative discussion rather than to establish broad participant agreement, we did not seek follow-up feedback from all Stage 1 participants.

Following initial ideation, the author team translated selected ideas into low-fidelity sketches and conceptual representations. These sketches were then discussed with M2, who commented on how the proposed feedback might be perceived during playing, whether the concepts aligned with existing practice routines, and what concerns they might raise, such as over-guidance, discomfort, or disruption of established technique habits. These discussions helped us refine the concepts and identify assumptions that required further caution. This bounded synthesis translated empirically identified breakdowns into grounded but still speculative design directions, sharpened through repeated participant feedback.

\section{Findings: Recurring Difficulties in String Learning for People with BLV (Stage 1) }
Our findings characterize recurring breakdowns in how violin and viola players with BLV learn and practice. Following Winograd and Flores \cite{winograd1986understanding}, we treat these breakdowns as moments that make otherwise taken-for-granted structures visible. Drawing on practice videos, lesson observations, and interviews with both students and instructors, we identify difficulties that emerge at the intersection of embodied technique, pedagogical mediation, and access to musical materials. Rather than isolated skill deficits, these challenges reflect structural mismatches between visually grounded string pedagogy and the non-visual conditions under which musicians with BLV learn.

Across interviews with string musicians with BLV and their instructors, three recurring difficulty areas emerged: right-hand (bowing), left-hand (hand-arm coordination, intonation), and score access/reading. These difficulties \textbf{Stage 1} presents these recurring difficulty areas as the empirical foundation for the design directions explored in \textbf{Stage 2}. We begin with Stage 1 below.

\subsection{Difficulties in Right-hand Control (Bow Hold and Bow Path)}
\label{sec:rhctrl}
\begin{figure}[h]
    \centering
    \includegraphics[width=1\linewidth]{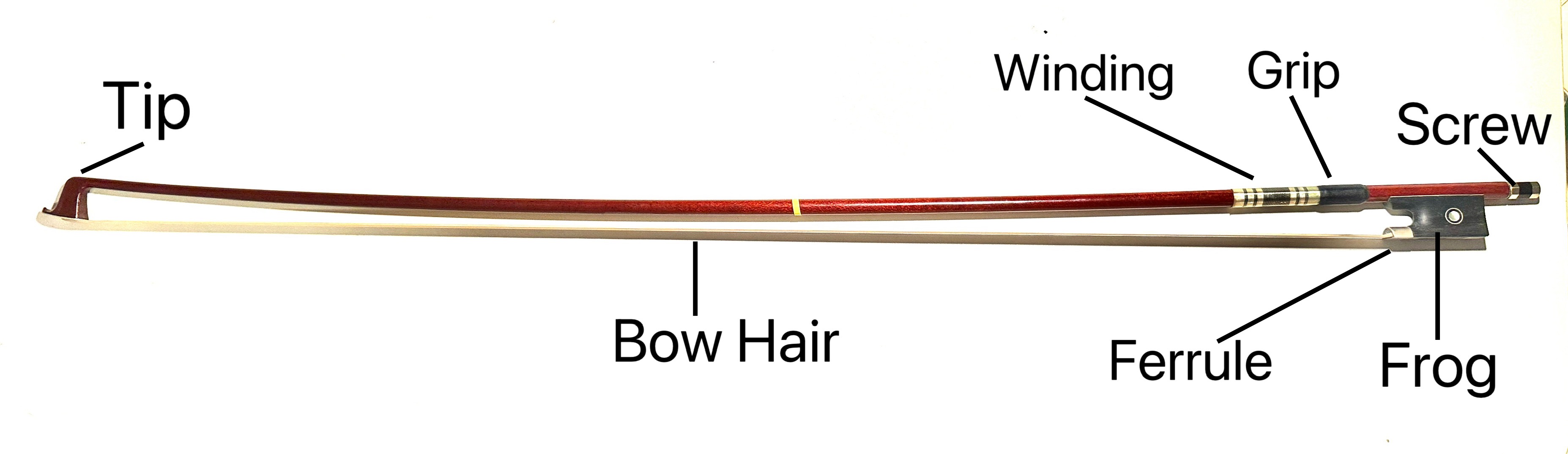}
    \caption{Bow anatomy}
    \label{fig:Bow}
    \Description{Labeled side view of a violin bow stretched horizontally. From right (player's hand) to left (far end), the labeled parts are: the screw at the back end of the frog, the grip (leather and winding wrapped around the stick where the hand sits), the frog (the black block that anchors the hair), the winding, the ferrule (the metal band holding the hair to the frog), the bow hair running the length of the stick, and the tip at the far end. The diagram is used to anchor subsequent descriptions of where learners with BLV lose spatial reference during bow control.}
\end{figure}

\begin{figure}[h]
    \centering
    \includegraphics[width=0.5\linewidth]{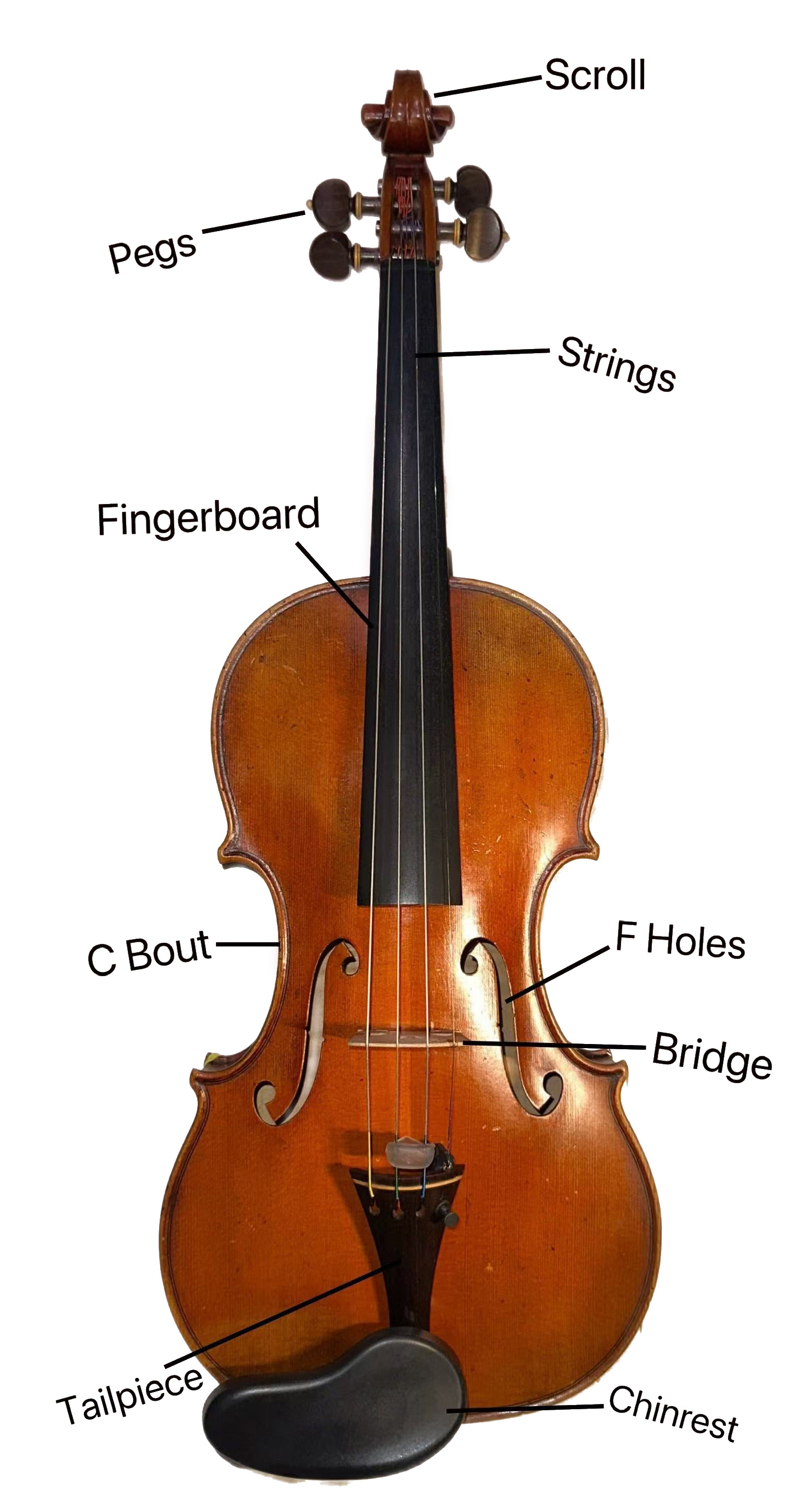}
    \caption{Violin anatomy}
    \label{fig:Violindiagram}
    \Description{Front view of a violin with labels indicating its main parts. From top to bottom: the curled \emph{scroll}; the \emph{pegs} that tune the four \emph{strings}; the \emph{fingerboard} running down the neck; the \emph{C-bout} (the curved inward waist of the body); the two f-shaped sound holes labeled \emph{F-holes}; the \emph{bridge} standing between the f-holes where the strings cross over; the \emph{tailpiece} anchoring the strings at the bottom; and the \emph{chinrest} attached beside the tailpiece. The diagram orients readers to the spatial relationships named throughout the findings.}
\end{figure}

All sampled string musicians with BLV and their instructors described right-hand bow control as a major challenge. In violin/viola learning, students usually rely on visual cues—watching the teacher and using a mirror—to self-correct two things: (1) bow hold and (2) bow path.

For bow hold, each right-hand finger has a specific role, and students must use minimal pressure to keep the hand loose and tension-free \cite{zhou2023right, auer1980violin, ihas2023whole, ihas2023applications}. For bow path, students try to keep the bow moving straight on the string, often taught as “bow stick roughly parallel to the bridge” across different parts of the bow. When the bow is not straight, tone often becomes scratchy and the bow may slide \cite{macleod2018teaching, hamann2009strategies}.
For learners with BLV, these visual checks are limited or unavailable, making right-hand errors harder to detect during self-practice. Participants described two linked problems. 

First, bow drift is difficult to detect in the moment. Because sound depends on steady friction between the bow hair and the string, small changes in direction or contact point can gradually lead to slipping, scraping, or unstable tone—often noticed only after the sound has already deteriorated. M2 described uncertainty about bow location and orientation, asking:\textit{“Where is my bow? Is my bow on the fingerboard, on my face, I have no idea”}, while I3 noted that \textit{“He (M2) can't see, he has to feel is this straight or not, this is very difficult”}. I3 also described the difficulty of giving spatial feedback: \textit{“sometimes when you say you have to go lower in the bow, he literally can’t see where the bow is, or where his bow is in relation to the string, and what part of the bow he’s in. The most he could do is feel where the extremities of his hands are”.} M4 similarily explained, \textit{“Playing the violin requires good bodily coordination. I can’t see whether the bow is straight or not—it’s related to spatial orientation and sense of direction. Building reliable muscle memory takes a very long time. It took me six or seven years before I stopped worrying about losing control of my right hand.”} Even M3, who had some usable vision, recalled: \textit{“right hand is hard, I spent a lot of time on getting better at the right hand skills, and bow straight was difficult when I was little.”}

Second, many participants described over-gripping at the frog as a protective strategy to avoid losing control. However, a strong grip increases tension in the hand, wrist, and forearm, reducing flexibility and fine control. The professor I1 talked about his experience in working with M1: \textit{“When he (M1) came for a lesson for the first time, I immediately noticed about his bow hold, the back of his hand was really tight, he was grabbing the bow very hard, and the arm was high up.”} Over time, this can limit tone development and cause injuries. I1 also noted that while they can often see drift and tension immediately, it is difficult to translate these visual diagnoses into verbal instructions that reliably produce the intended adjustment for M1, \textit{“He (M1) doesn't understand when I say your bow hold is tight, your bow is not straight, I need to grab his hand and make him feel, or I play, and let him touch my hand.”}

\subsection{Difficulties in Left-Hand Control (Arm Angle and Intonation)}
\label{sec:lhctrl}
All sampled string musicians with BLV and their instructors also described left-hand control as a major challenge, especially wrist alignment, arm rotation/angle, and resulting intonation issues. In upper-string pedagogy (violin/viola), students are typically taught to keep the left-hand shape stable—often described as curved fingers with a firm fingertip contact—while maintaining a relatively neutral wrist (hand–wrist–forearm aligned). Beginning players often bend the wrist outward or collapse it inward (sometimes called a “pizza wrist”). Sighted beginners can usually detect this quickly by watching themselves in a mirror, but participants reported that they often cannot feel whether their wrist is straight during playing and only learn about the issue when a teacher or another person observes them.  I1 noted: \textit{“I have a mirror in my studio, and my students can see their posture, but what about M1, he can't see.”} A persistently bent wrist can have several consequences if not addressed early. It can increase tension in the left hand and forearm, reduce agility for fast passages, and interfere with more advanced techniques. It may also contribute to strain or injury over time.

Instructors also emphasized intonation as a recurring instructional focus. Accurate intonation depends not only on finger placement, but also on subtle left-arm angle and rotation, which help the fingers land consistently across strings. When the arm angle is off, the wrist often becomes misaligned, and pitch accuracy suffers. Instructors further noted that left-arm positioning affects the overall instrument setup in space, which can block progress on other technique areas. As I3 explained, \textit{“he (M2) was holding the viola really high up…there were huge problems with his right arm that just could not be addressed unless we move the instrument lower”.}  Because string learners with BLV have limited access to visual self-monitoring of arm, wrist, and instrument position, these alignment-related problems can be harder to diagnose and correct during self-practice.

\subsection{Score-Related Challenges (Score Learning and Daily Practice Assistant)}
In our data, score study was difficult not only because access is limited, but because the available formats do not support real-time playing and detailed interpretation (e.g., bowings, fingerings, phrasing). This pushes learners toward memorization and listening-based learning, and shifts more note-clarification work into lessons.

Score study has been a major struggle for all of the learners with BLV. Many described early music learning as dependent on a sighted helper: M4 said \textit{“Even if it’s just scales, I need to find someone who can see, and read the notes to me”.} They also explained a basic mismatch between reading and playing: M2 explained \textit{“There’s an issue, braille is read by touching, I can’t play when I’m touching the braille”}, M1 also supports this using an example: \textit{“just imagine when you play in an orchestra, conductor said open page 29 and find a bar, okay, I don't have a third hand, I can't play while finding the bar.”} As a result, memorization becomes the default pathway. M4 said \textit{“So I memorized all my music, from scale to etudes, concerto, and sonata”.} Over time, M4 reported shifting toward ear-based strategies:\textit{“Slowly, my ear got better, so I listen first, I sometimes slow down the recording, and if I can’t hear clearly, I ask the teacher during my lesson.”} 

While music software can provide some accessibility functions, participants emphasized that these tools often remain insufficient for their needs, especially when information is delivered slowly (e.g., one note at a time) and require continuous manual control. More importantly, participants noted that the tools do not capture many performance-critical details: M2 mentioned that\textit{“It tells me in very little detail, like the bowing and fingering, like some special techniques”.}  For example, M2 explained \textit{“The MuseScore only tells me the length of the note. I don’t need it to tell me, I already know the melody, I already have the recordings on YouTube”.} For M3, who can see the score, he also mentioned that \textit{“everything I read needs to be enlarged by 125\%, this is very important especially when I read the conductor's score, cause the notes are much smaller.”}

Instructors reinforced this gap by describing how accessible notation tools can help with basic pitch and rhythm, but often break down in musically complex contexts. I2 explained:\textit{“You can obviously check in with MuseScore to get the note values and pitches…but one of the things that’s very difficult in the Tchaikovsky is, there’s a lot of rubato, I’m so used to hearing that, but the rubato on the recordings were very confusing to him (M2), he was getting confused with what the rhythm was, or how the notes were grouped”.}  

For fast passages, I2 described an additional barrier: \textit{“there are all those fast runs where one or two notes change each time, he (M2) gets it, but it’s hard to go back and work on these”,} and \textit{“it’s been a struggle for him (M2) to find the patience to work on small bits of music”.} As I2 summarized: \textit{“I think it just needs time, but time is in such short supply… I have not found a great way to deal with that yet”.}  Instructors also noted that note-related problems often spill into lesson time and require ad hoc workarounds. \textit{“when it’s obvious a more difficult fix, I would just stop in the lesson and make a 20-second recording for him so he could access that easily and not have to go back through all his files”.} (I2)

Finally, because all student participants were international students who studied in one country and then moved to another, several described language barriers not only in speaking and writing, but also in relearning braille conventions after moving, since literacy from their home country's system did not fully transfer—adding to their existing access burden. M1 said: \textit{“I learned Chinese braille when I was in middle school, but now, I don't use that anymore.”}

\section{Findings: Speculative Design Ideation (Stage 2)}
Building on the empirical findings from Stage 1, we developed a set of speculative design concepts to explore how recurring breakdowns in string learning for people with BLV might be addressed through interactive systems. Rather than proposing deployable solutions, this stage adopts a Research through Design perspective to translate observed difficulties into design questions and directions. The goal is to surface alternative ways of supporting embodied learning and independent practice when visual reference and continuous instructor feedback are unavailable.

Across the empirical study, participants repeatedly described challenges related to sensing technique states in real time, coordinating fine-grained bodily movements, and integrating feedback into hands-busy practice routines. In conventional string pedagogy, these processes are heavily supported by visual demonstrations, mirrors, and immediate visual diagnosis by teachers. For learners with BLV, the absence of such visual scaffolding creates gaps that are only partially addressed by existing tools, which often rely on physical constraint, episodic correction, or post-hoc feedback. Instructors noted that such issues can persist unnoticed into advanced stages of playing unless explicitly addressed. I1 reflected that \textit{“sighted students often catch wrist collapse early by looking in a mirror, but for blind students the problem can stay unnoticed for a long time.” }

The speculative concepts presented below respond to these gaps by exploring designs that externalize otherwise invisible technique states—such as bow trajectory, grip tension, wrist alignment, and practice context—through non-visual, low-attention feedback. Each concept is grounded in specific empirical breakdowns identified in Stage 1 and reflects an intentional shift from corrective or restrictive aids toward perceptual augmentation and practice scaffolding. Together, these concepts articulate a speculative design space for supporting string learning for people with BLV that could complement, rather than replace, pedagogical instruction.

\subsection{Augmented Bow Track}
\subsubsection{Design inspiration.} \label{5.1.1}
Section \ref{sec:rhctrl} shows that right-hand control remains difficult for learners with BLV because key aspects of bowing---such as keeping a stable contact point and a straight bow path---are conventionally learned through visual reference (teacher demonstration and mirrors). Without an in-the-moment spatial reference, learners often notice bow drift only after tone quality has already deteriorated, and some adopt protective strategies such as over-gripping at the frog, which increases tension and reduces fine control. I1 seconded this: \textit{“When I say dynamics change to louder, you have to move back towards the bridge. String instruments has to be trained, if you don't work on that area of counterpoint which is near the bridge and at the beginning, the instrument will get confused, and produce ugly sound.”} Instructors can frequently \emph{see} bow drift and tension quickly, yet translating these visual diagnoses into precise verbal cues that reliably produce the intended adjustment is challenging for learners with BLV and related teaching. 

Several participants described using low-tech pedagogical aids to compensate for the lack of visual feedback. For example, M1 recalled how his parents used rubber bands and a pencil attached near the bridge to indicate whether the bow deviated from a straight path: \textit{“My parents just helped me to fix the problem, and they didn’t use a mirror. They just used the rubber band and pencil, just tell me what is straight play, how to play straight... When you play, if you crash the rubber band and the pencil, you understand it’s not straight.”} 

Similar strategies were also adopted in formal instruction. I3, who served as an assistant instructor working alongside I1 in teaching M1, described how rubber bands were incorporated into their lessons: \textit{“He (M1) had the pinky house and the rubber band before, then we removed the pinky house, and have been using the rubber band till now, it helps his middle two fingers stay down at a certain level.”} 

I1 still emphasized their limitations and noted that the feedback they provided was binary and potentially disruptive, and that correct bow trajectory alone did not guarantee correct arm use: \textit{“Yet, the rubber band is effective, but you still need your teacher to teach you how to open your elbow. Even if you play straight, your elbow can still be wrong.”}  These accounts highlight how existing low-tech aids rely on physical obstruction to signal error and remain dependent on close teacher supervision.

\begin{figure} [h]
    \centering
    \includegraphics[width=1\linewidth]{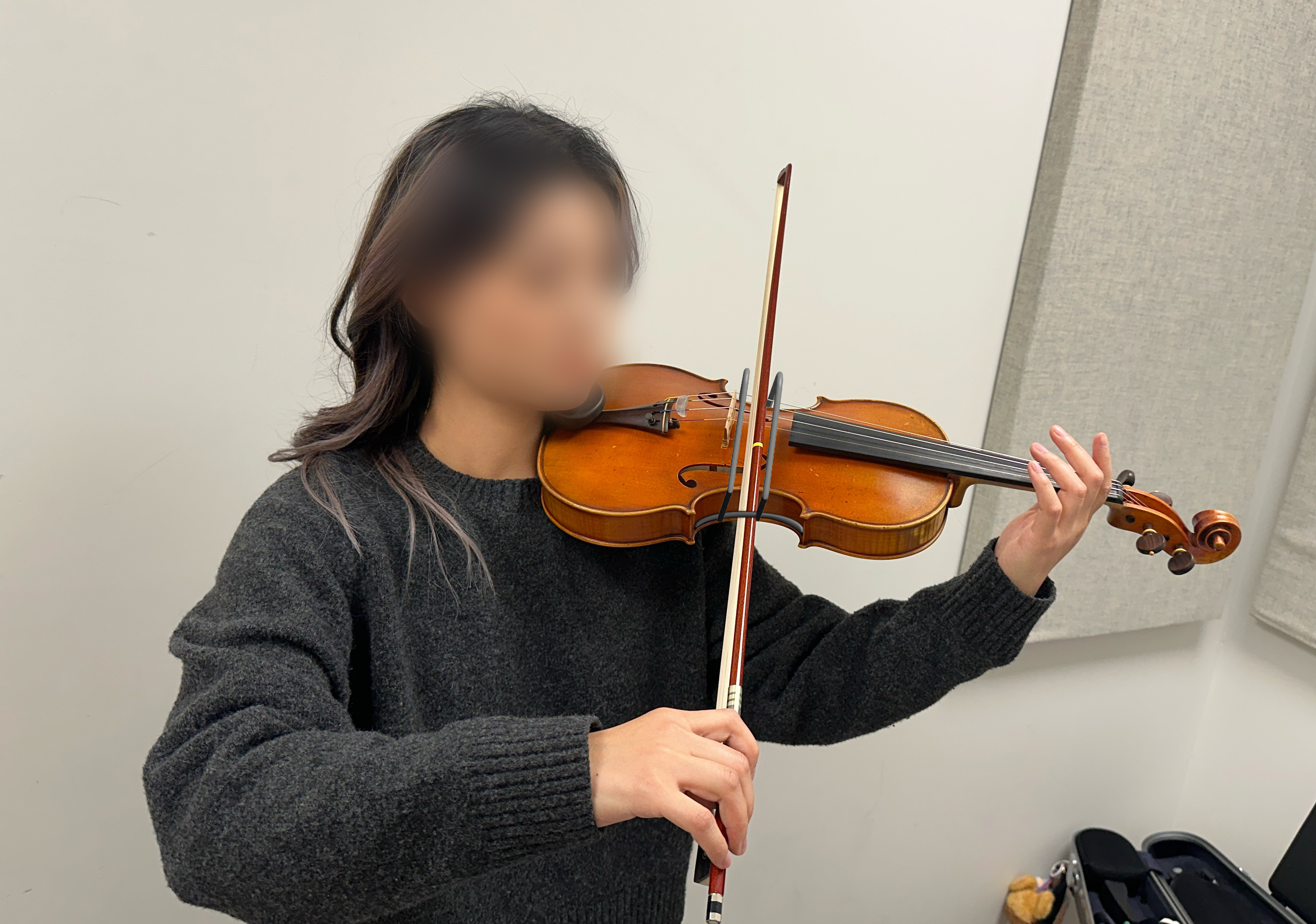}
    \caption{Bow track}
    \Description{Photograph of a string player, face blurred for anonymity, seated in playing position with a viola under the chin and a bow across the strings. The bow is being held mid-stroke across the instrument, with the right arm partially extended. The photograph motivates the design concept described in \S5.1 by grounding the discussion of bow trajectory in a real practice posture.}
    \label{fig:bowtracker1}
\end{figure}
\begin{figure} [H]
    \centering
    \includegraphics[width=1\linewidth]{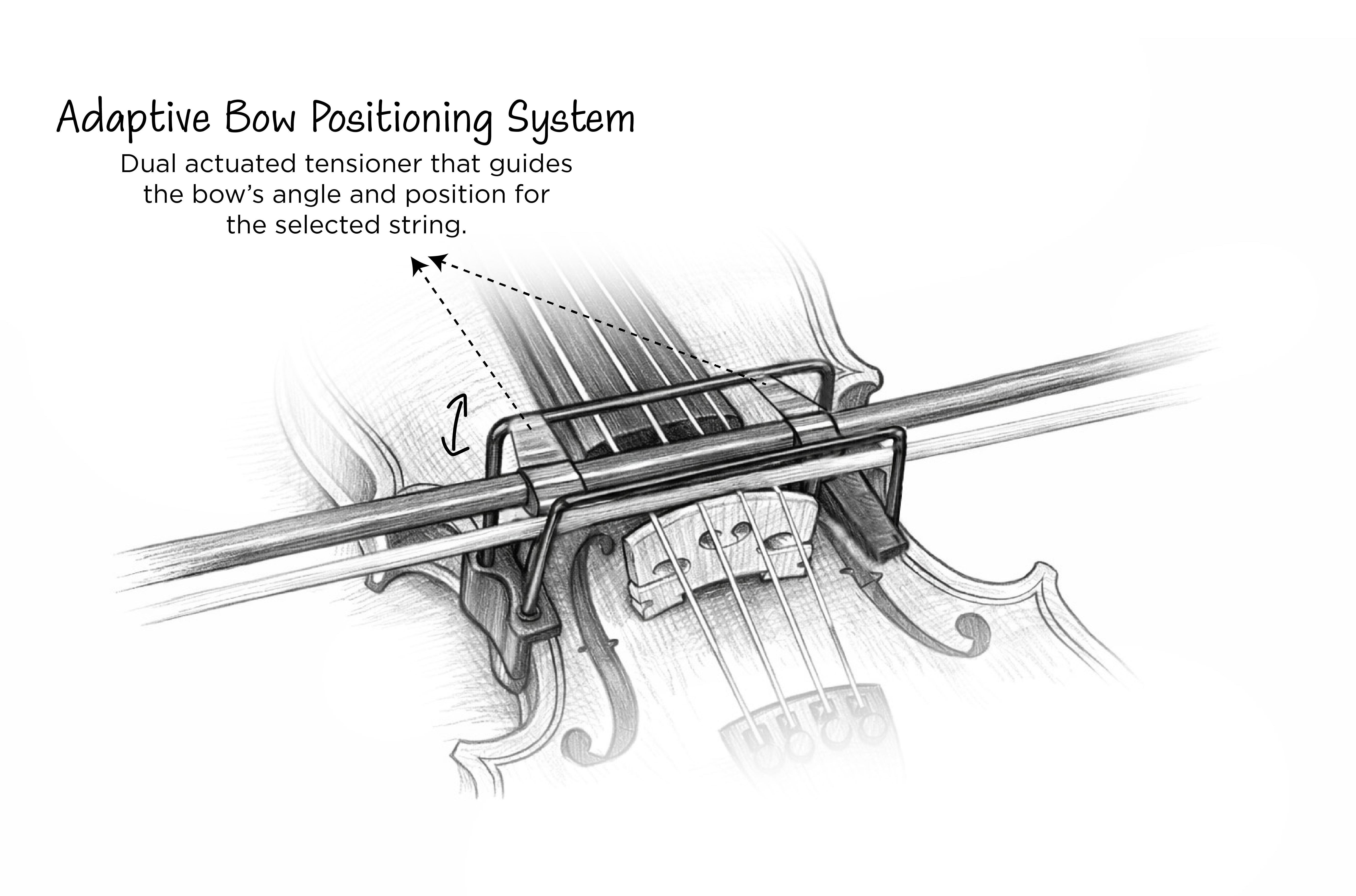}
    \caption{Augmented bow track that uses friction as non-visual feedback for students with BLV}
    \Description{Hand-drawn sketch of the proposed \emph{Adaptive Bow Positioning System} clipped onto the violin between the bridge and the fingerboard. A dual actuated tensioner bridges the strings and rests lightly against the bow as it travels across the string. A caption on the sketch reads ``Dual actuated tensioner that guides the bow's angle and position for the selected string.'' The tensioner provides friction that increases as the bow drifts laterally from the straight-path position, giving players with BLV continuous non-visual feedback on bow trajectory without fixing a rigid path.}
    \label{fig:placeholder}
\end{figure}

\subsubsection{Design concept.} Inspired by these strategies, we conceptualized a \textit{bow tracker augmentation} to the traditional bow holder. As shown in Figure \ref{fig:bowtracker1}, it has a clip-on add-on positioned between the bridge and fingerboard, using two moving tensioners that hold the bow and shift position according to the string being played. Rather than enforcing a fixed path or posture, the tracker is designed to make bow deviation perceptible through graded friction—the further the bow drifts from the optimal position, the more resistance the player feels—with the goal of building muscle memory for correct position through repeated practice. The system targets spatial awareness of bow motion, a core yet visually taught aspect of string pedagogy, while preserving freedom of movement.

We envision the bow tracker supporting right-hand learning through four core functions:

\begin{itemize}
    \item \textbf{Non-visual feedback for bow drift and contact-point change.} The augmentation is designed to mechanically signal lateral deviation of the bow relative to the intended straight path (e.g., drifting toward the fingerboard or toward the bridge) and changes in the contact point that often precede slipping, scraping, or unstable tone.
    \item \textbf{Continuous, low-attention feedback.} Instead of a binary “collision” signal, the system is intended to provide graded non-visual feedback in the form of friction that increases as deviation grows, enabling learners to correct mid-stroke rather than after tone degrades.
    \item \textbf{Adaptive bow-lane awareness for dynamics and practice.} Bow lanes—playing closer to the bridge or fingerboard to shape dynamics and timbre—are particularly difficult for learners with BLV to perceive. The tracker can support awareness of different bow lanes by dynamically adjusting sensitivity regions based on practice goals or musical context. In a practice mode, learners can intentionally explore different lanes for specific passages, building an embodied understanding of how right-hand position and arm movement relate to sound production, without relying on visual cues.
    \item \textbf{Practice scaffolding and reflection.} The tracker can be used to structure practice into targeted drills (e.g., slow open strings, specific bow regions) and optionally provide post-practice summaries (e.g., frequency and direction of drift) to support teacher--student debriefing.
\end{itemize}

This bow tracker augmentation is not intended to define a single “correct” technique or replace instruction. Instead, we envision it functioning as a learning companion that could externalize an otherwise invisible spatial relationship, supporting violinists with BLV in developing embodied understanding through repeated practice. By shifting from restrictive guidance toward perceptual augmentation, the design responds directly to participants' lived strategies while attempting to address their limitations; whether it does so in practice remains to be evaluated.


\subsubsection{Feedback from a musician with BLV.}
M2 responded positively to the proposed concept, describing the initial bow-holder idea as \textit{“very feasible” }and rating it as \textit{“an 8/10.”} He emphasized that \textit{“using touch for feedback is better,”} explaining that \textit{“using sound to signal feedback is distracting for beginners.”} From his perspective, \textit{“it’s more feasible to just augment the existing bow holder,” }with friction serving as a preferable feedback modality.

M2 further suggested constraining the wooden part of the bow with a soft material,\textit{ “like siliconegel,”} which would make it \textit{“harder to damage the soft part of the bow”} while also making it\textit{ “harder to get the bow fully stuck.” }He emphasized that feedback should scale with performance error, proposing to \textit{“have friction in proportion to the error in angle for the bow, to build a positive memory when having a correct angle.”}

Finally, M2 expressed a clear preference for a multi-point guidance structure, stating that \textit{“two point contact is better than one.” }In his view, this configuration not only \textit{“can straight the bow,”} but also\textit{ “can form good habit of not touching other strings,” contributing to “a cleaner tone.”}

Together, M2’s feedback affirms the overall promise of the concept while underscoring the importance of tactile, graded, and non-restrictive feedback that supports habit formation without over-constraining movement. 

\subsection{Augmented Bow Hold Aid}
\subsubsection{Design inspiration.}
Section \ref{sec:rhctrl} and \autoref{5.1.1} also reveals that over-gripping and fine control also faces similar challenges as the player typically relies on visual reference from instructors or instructional materials.

\begin{figure} [H]
    \centering
    \includegraphics[width=0.7\linewidth]{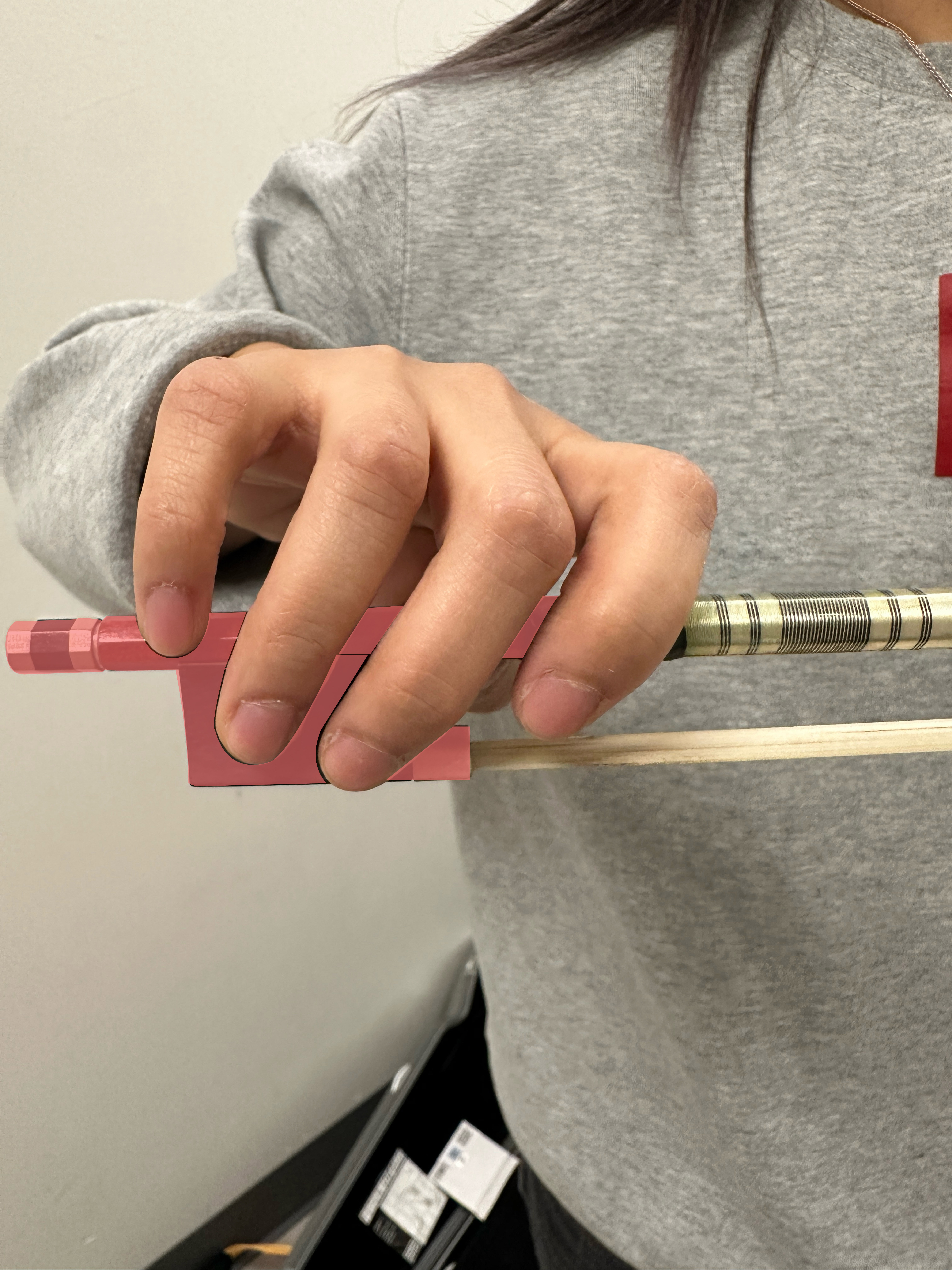}
    \caption{Pinky holder that wraps around the bow's holder}
    \Description{Close-up photograph of a right hand holding a bow at the frog. A small molded pinky holder is attached to the bow stick and wraps around the region where the holding fingers contact the stick. The photo shows the finger placement the augmented version is intended to preserve.}
    \label{fig:pinky1}
\end{figure}

\begin{figure} [H]
    \centering
    \includegraphics[width=1\linewidth]{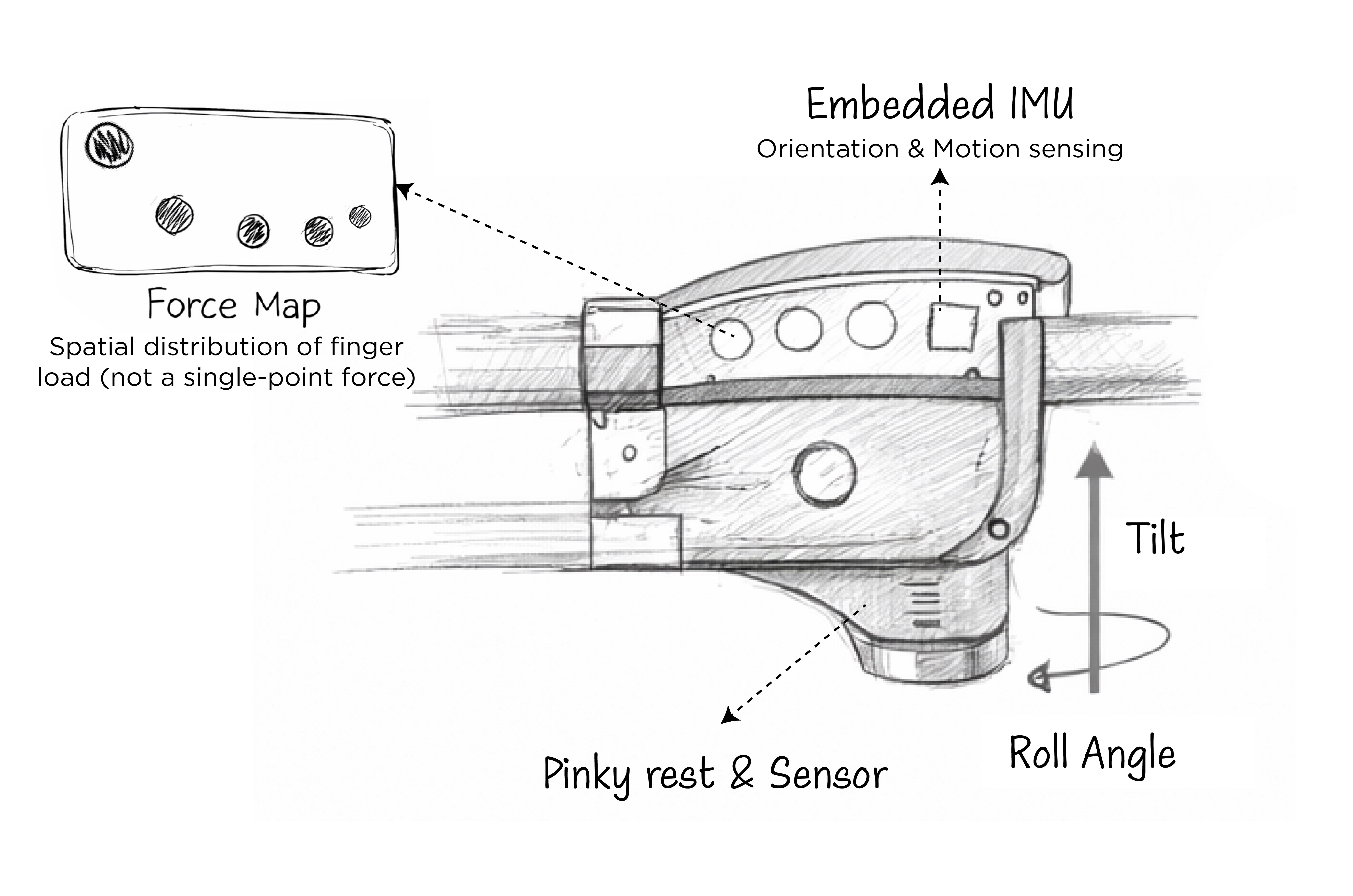}
    \caption{An augmented pinky holder that detects and maps the pressure of the holding fingers, as well as sense if the bow is moving in the correct orientation}
    \Description{Exploded-view sketch of the proposed augmented pinky holder, based on a conventional bow frog enlarged slightly to embed sensors. The sketch labels four sensing components. An \emph{embedded IMU} on top of the frog measures orientation and motion (tilt and roll angle). A \emph{force map} on the holding surface captures the spatial distribution of finger load rather than a single-point force. A \emph{pinky rest \& sensor} sits under the pinky to detect whether the finger is actually resting in position. Arrows annotate tilt and roll axes. Together the components infer both grip pressure distribution and bow orientation during playing, supporting real-time feedback about where and when over-gripping occurs.}
    \label{fig:pinky2}
\end{figure}

\subsubsection{Design concept.} Based on this need, we conceptualized \textit{an augmented bow hold aid} as a bow-mounted augmentation (Figure \ref{fig:pinky2}) intended to support awareness of the bow hold, grip tension, and load distribution. We preserved the overall geometry of a traditional frog for compatibility with existing bows and minimal disruption to established motor habits, drawing on pinky holds (see Figure \ref{fig:BLV RH tools}), which are already widely employed in beginner string pedagogy to correct the placement and load distribution of the pinky finger \cite{topper2002correcting}. The device would slightly enlarge the grip volume compared to a standard frog to accommodate embedded sensing components; this is a real tradeoff, since prolonged use may lead players to adapt to a larger grip and feel discomfort returning to an unaugmented bow, a limitation we revisit later in the paper.
This design would use thin-film pressure sensors (e.g., Velostat) to produce a spatial \textit{force map} across multiple contact points; as shown in Figure \ref{fig:pinky1}, the highlighted red areas correspond to sensor-instrumented regions where right-hand finger pressure is measured. By mapping where pressure concentrates and how it shifts over time, the system could infer both finger placement and over-gripping patterns (e.g., excessive force near the frog or at specific finger pads). A compact inertial measurement unit (IMU), similar to those in smartphones, could capture bow orientation and kinematics; combined with the force map, this could allow the system to localize \textit{where} and \textit{when} grip pressure increases as bow angle or stroke changes, offering evidence of tension beyond tone alone. With the support of modern AI, such a system could serve real-time, low-attention feedback that reinforces correct bow hold when students with BLV practice without direct instructor supervision, indicating \textit{which contact region is too tense} and suggesting \textit{which finger to relax or reposition} (e.g., ``reduce index pressure,'' ``shift load toward the pinky,'' or ``soften thumb contact''), rather than only flagging error.

We envision the augmented pinky holder supporting habit-building of correct bow hold by:
\begin{itemize}
    \item \textbf{Ergonomic shape.} The specific shape of the design is intended to guide the fingers to be in the correct position and encourage a balanced load path that reduces compensatory gripping.
    \item \textbf{Force mapping based feedback.} The holder senses each finger's position and force, is intended to sense each finger's position and force and provide real-time feedback on how to adjust the bow hold. Specifically, it visualizes/encodes a multi-point force map and highlights hotspots, enabling targeted relaxation (“loosen here”) and micro-adjustments (“shift contact slightly”) based on where pressure is excessive.
    \item \textbf{Bow angle and movement.} By sensing the orientation and kinematics of the bow, the holder is also intended to continuously provide feedback on bow angle and stroke consistency, and \textbf{relates these changes to pressure redistribution} (e.g., increased tilt correlating with a spike in index pressure), helping learners maintain a relaxed hold while keeping stable bow control.
\end{itemize}

\subsubsection{Feedback from a musician with BLV.}
M2 generally responded positively to the concept while raising concrete concerns about posture transfer and long-term use. M2 described the idea as promising, stating that \textit{“the idea is good” }and that he thought \textit{“it can be useful.”} At the same time, he cautioned that the hand posture with and without the pinky holder \textit{“is not the same,” }suggesting that the aid \textit{“might not be that helpful when you take it away.”} Drawing on prior experience, M2 noted that attaching materials to the bow grip altered tactile perception, explaining that \textit{“I have tried having rubber attachment to the bow’s holding place… It resulted in different feeling.”} He further expressed concern that embedding sensors could \textit{“deform the correct posture.”}

Together, these responses underscore a design tension for the augmented bow hold aid: while participants recognized its potential value for awareness-building and early-stage learning, they also emphasized risks related to posture transfer, sensory distortion, and over-reliance on grip augmentation. These concerns informed our framing of the device as a temporary, practice-oriented scaffold rather than a permanent replacement for conventional bow hold training.

\subsection{Left-Hand Alignment Wearable}
\subsubsection{Design inspiration.}
Based on the challenges described in Section \ref{sec:lhctrl}, both participants with BLV and instructors consistently emphasized left-hand control as a persistent difficulty, particularly wrist alignment, arm rotation, and their impact on intonation. Participants described being unable to reliably sense whether their wrist and forearm remained aligned during playing, often discovering misalignment only when a teacher or another person observed them. As M1 explained, \textit{“Even now, I often don’t know whether my wrist is straight when I’m playing. I usually only realize there’s a problem when my teacher or someone else watches me and tells me.”}


These findings motivated us to explore a wearable approach that could make “invisible” alignment and rotation more perceptible to learners with BLV during self-practice. Our design inspiration came from two converging observations. First, teachers in our study frequently used tactile and hands-on instruction—physically repositioning the student’s arm, wrist, or fingers—to “show” what a correct alignment feels like, suggesting that left-hand technique is fundamentally embodied and difficult to convey through words alone. Second, participants repeatedly described relying on proprioception and muscle memory, yet also acknowledged that proprioception can be unreliable when the body gradually adapts to misalignment. This tension pointed to an opportunity for a device that does not prescribe a single correct posture, but instead helps learners notice when they have drifted away from a baseline alignment. We also drew inspiration from commercially available neuro/fitness wearables (e.g., the Meta-style “neuro band” concept) that sense muscular activity and translate subtle movement patterns into feedback, suggesting a plausible design direction for tracking small changes in arm rotation and wrist posture without requiring vision-based capture.

\begin{figure} [h]
    \centering
    \includegraphics[width=1\linewidth]{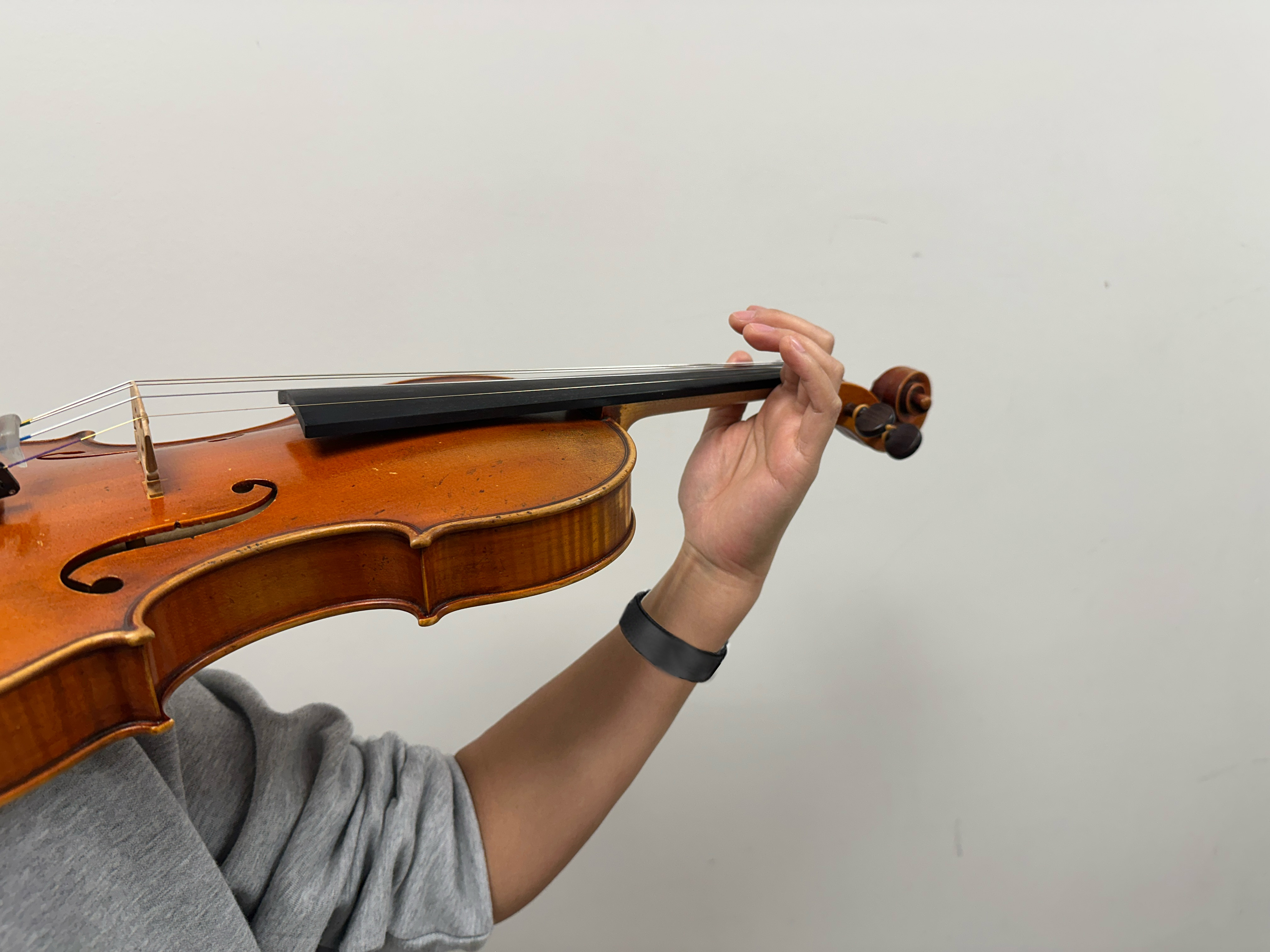}
    \caption{Haptic wristband for left hand positioning}
    \Description{Photograph of a string player's left arm in playing position. The violin is held up under the chin; the left hand is on the fingerboard. A thin black wristband, roughly 1.5\,cm wide, is worn on the lower forearm just above the wrist, illustrating the form factor and placement of the proposed left-hand alignment wearable.}
    \label{fig:wearable2}
\end{figure}

\begin{figure}[h]
    \centering
    \includegraphics[width=1\linewidth]{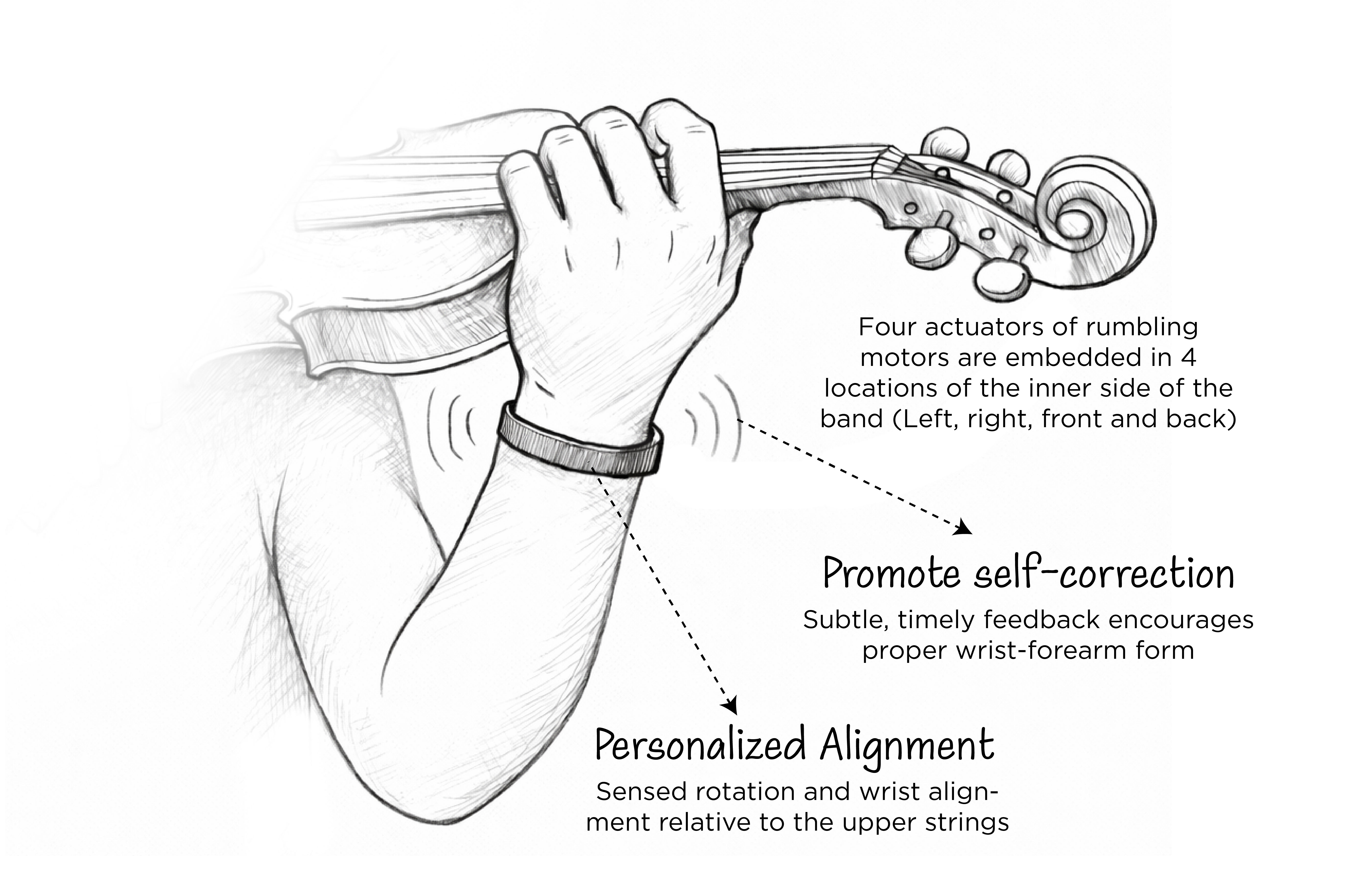}
    \caption{A wrist band that detects the left hand orientation and gives haptic feedback to remind the student to correct posture}
    \Description{Hand-drawn sketch of the proposed left-hand alignment wristband in use. A player's left hand is shown on the violin neck in playing position. The band is worn on the forearm, with callouts indicating two functions. ``Personalized Alignment'' (callout on the band): the device senses rotation and wrist alignment relative to the upper strings, anchored by the scroll connection. ``Promote self-correction''(callout on the opposite side): four actuators of rumbling motors embedded at four locations on the inner side of the band (left, right, front, back) deliver subtle, timely haptic feedback that encourages proper wrist-forearm form.}
    \label{fig:wearable}
\end{figure}

\subsubsection{Design concept.}
We propose a lightweight left-hand wearable that supports string musicians with BLV in calibrating wrist–forearm alignment and arm rotation while fingering. As sketched in Figure \ref{fig:wearable}, the device is a narrow band (approximately 1.5 cm wide) worn around the lower forearm near the wrist or slightly above it, where rotation and wrist deviation tends to manifest as changes in muscle activation and surface movement. The form factor is intentionally minimal to avoid interfering with bow hold, shifting, or vibrato, and to remain compatible with common practice routines. The attachment is designed to be lightweight and non-obstructive, preserving left-hand mobility while maintaining a repeatable reference across sessions. The wearable’s core function is to detect patterns associated with wrist collapse and insufficient or excessive forearm rotation when moving across strings, then provide subtle feedback that supports self-correction. As discussed further in Section 6, reliably inferring this from a narrow wristband remains an open sensing question rather than a solved one.


We envision the left-hand wearable supporting string learning with BLV through four core functions:

\begin{itemize}
    \item \textbf{Personalized alignment sensing.}
The wearable is intended to sense wrist alignment and forearm rotation using common wearable sensing modalities (e.g., orientation and surface movement), learning an individualized baseline rather than enforcing a universal “correct” posture. Anchored by the scroll connection, sensed rotation and deviation can be interpreted relative to the instrument, enabling more consistent detection during string crossings.

    \item \textbf{Event-based feedback for misalignment.}
When sustained deviation from the personalized baseline is detected: such as wrist collapse or insufficient arm rotation, the band is intended to provide brief, localized feedback (e.g., haptic pulses), avoiding continuous or distracting instruction.

    \item \textbf{Support for intonation-related technique.}
By stabilizing wrist–forearm alignment and string-specific arm rotation, the device is intended to indirectly supports more reliable intonation without attempting to correct pitch directly through audio matching.

    \item \textbf{Scaffolding for lessons and self-practice.}
The wearable can be used selectively during warm-ups or technical drills to extend teachers’ hands-on guidance into self-practice, and removed during musical interpretation to avoid over-reliance.
\end{itemize}

\subsubsection{Feedback from a musician with BLV.}
M2 responded positively to the left-hand wearable’s potential to provide timely posture awareness during practice. M2 emphasized the value of real-time feedback, noting that “it has a good real-time feedback” and that it “saves time while fixing your posture.” He contrasted this with delayed correction, explaining that “after you reach a more advanced level then notice this mistake, it’s very hard to fix it,” highlighting the importance of early intervention for left-hand alignment issues.

At the same time, M2 raised questions about feedback modality and sensing reliability. He suggested that \textit{“it would be better if it uses beep… beep… sound feedback,”} proposing audio cues as a way to signal misalignment. This prompted reflection on earlier concerns raised in the study, as A2 immediately responded, \textit{“Didn’t you mention in the previous interview that audio feedback is distracting?”} M2 acknowledged this tension, replying, “Right,” and reframed his concern as one of sensing robustness rather than modality alone.

Specifically, M2 questioned whether a wristband form factor could reliably capture wrist posture across placements, explaining that\textit{ “the wristband is hard to monitor the wrist depending on where exactly you put it.”} He proposed an alternative configuration that might better align with anatomical movement, suggesting that \textit{“maybe it doesn’t need to be a wrist band,”} and that it could instead be \textit{“a few tape-felt stripes on the back of the hand,”} extending “from the back of the hand to the forearm.” This suggestion reflects a desire for a more distributed and placement-tolerant sensing surface, rather than a rigidly localized wearable.

Together, these responses highlight both the perceived usefulness of timely, alignment-focused feedback and ongoing questions about feedback modality and form factor. Participants valued the wearable’s potential to surface otherwise invisible posture drift, while also emphasizing the need for designs that balance sensing reliability, attentional load, and flexibility in how the device is worn.

The three design directions represent early-stage explorations grounded in Stage 1 findings. M2's feedback represents a single perspective. Pedagogical alignment with established practice and classroom integration remain unvalidated and require further development before deployment.

\section{Discussion}
Returning to our two research questions: Stage 1 (Section 4) addresses RQ1 by identifying three interconnected difficulty areas that recur across right-hand bow control, left-hand alignment and intonation, and score-related practice work during self-directed learning. Stage 2 (Section 5) addresses RQ2 by translating these breakdowns into three speculative design directions—the Augmented Bow Track, the Augmented Bow Hold Aid, and the Left-Hand Alignment Wearable, that explore how non-visual, practice-compatible calibration might be supported during independent practice. The remainder of this section interprets these findings through a design-oriented lens, situating string learning for BLV musicians within broader questions of embodied interaction, accessibility, and skill acquisition. Beyond music learning, our findings suggest that accessibility challenges in embodied domains are often not only about accessing instructions or representations, but about accessing ongoing perceptual reference during action. From this perspective, centering musicians with BLV’s lived experiences enables new ways of conceptualizing non-visual feedback, independent practice, and disability-centered design knowledge \cite{draper2022music}, while also reframing accessibility as a problem of embodied calibration rather than information access alone.

\subsection{Designing for Musicians with Blindness and Low-Vision: Making Technique States Perceivable Without Vision}

Designing for musicians with BLV is often framed as a problem of information access: translating visual materials such as scores, demonstrations, or diagrams into alternative formats \cite{lu2023there,zhang2025accessibility}. Our findings reveal a distinct barrier of the loss of \textit{dynamic perceptual reference} during action. In upper-string learning, many of the most consequential aspects of technique are continuous, embodied states—bow trajectory and contact point, distributed grip tension in the right hand, wrist–forearm alignment and arm rotation in the left hand—that sighted learners routinely monitor through mirrors, visual comparison with instructors, or watching the bow relative to the bridge. For learners with BLV, these states remain largely invisible during self-practice.

Designing for musicians with BLV therefore needs to shift from substituting visual representations to externalizing otherwise invisible technique states into forms that can be perceived non-visually and acted upon in real time—moving accessibility beyond post-hoc translation toward new perceptual reference systems for embodied action. Rather than describing what a specific technique \textit{“should look like,”} effective designs could support learners in sensing where the bow is traveling, how pressure is distributed across the hand, or when alignment has drifted, while movement is unfolding.

Importantly, our study shows that these technique states are not binary errors to be corrected, but graded and context-dependent ranges that shift with musical goals, repertoire, and individual physiology. The design concepts presented in this work exemplify this by amplifying awareness of spatial drift, tension hotspots, and alignment changes while preserving freedom of movement and expressive variability—early probes of how learning might be supported without prescribing a single correct posture, collapsing technique into rigid constraints, or replacing teachers’ judgment. Their pedagogical integration and practical value remain to be evaluated.

By centering musicians with BLV, this work surfaces the visual assumptions embedded in embodied pedagogy—assumptions that often remain invisible when designing for sighted users—and shows how accessibility challenges in such domains often concern perceptual reference during action, not information access alone \cite{moffatt2004designing,howey1995technology,steriadis2003designing}, illustrating how disability-centered design can expose foundational interactional dependencies and open new design spaces rather than merely extending existing systems to marginalized users. These interactional dependencies are not experienced evenly across learning contexts: when access to specialized instruction, adaptive resources, and sustained pedagogical support is limited, the consequences of relying on visually grounded teaching practices become even more pronounced.

This reframing is particularly consequential in contexts where pathways for individuals with BLV to pursue professional or advanced musical training are severely constrained, including many Global South settings \cite{ran2025users,ran2025understanding}. Several participants, including M1 and M2, described becoming a musician as an exceptional trajectory within their local contexts, where blind individuals were often steered toward a narrow set of other occupations regardless of interest or talent, and specialized instruction remained concentrated in a few elite institutions. In such contexts, independent practice tools matter even more, since continuous instructional support cannot be assumed.

The three design directions also sit at different distances from current technological feasibility, and we think this variation is itself informative. The Augmented Bow Track and the Augmented Bow Hold Aid draw on sensing and actuation approaches largely available today—thin-film pressure sensors, compact IMUs, and simple actuated tensioners—so they are closer to near-term realizable prototypes, pending engineering and safety refinement. The Left-Hand Alignment Wearable raises a more open question: as M2 noted, a narrow wristband's ability to reliably infer wrist–forearm alignment depends heavily on placement and individual anatomy, an open problem in wearable sensing rather than a disqualifying one—so we read it as a further-out speculative direction than the other two. Even where a specific form factor may not yet support the intended sensing, the underlying design commitments—graded rather than binary feedback, touch rather than sound, and an individualized rather than universal baseline—are the design knowledge we would want future work to carry forward regardless of which exact sensing mechanism is eventually used.

\subsection{Design Implications}
\textbf{First, effective designs could make key technique states perceivable during hands-busy practice.} String learning depends on continuous, fine-grained calibration of body position and movement. For violin/viola players with BLV, this is harder to develop not because of skill deficits, but because conventional pedagogy assumes fast visual self-checking (mirrors, teacher observation). Tools could externalize key technique states during self-practice: right-hand supports could provide feedback on grip pressure and bow drift, allowing mid-stroke adjustment; left-hand supports could signal wrist-forearm drift and arm rotation without restricting movement; score work tools could support the full learning workflow (navigation, verification, correction) when both hands are occupied. These implications remain exploratory: the concept directions presented here have received feedback from one musician with BLV (M2) but have not been evaluated with instructors or tested in deployed contexts, and further research is needed to assess pedagogical alignment, movement freedom, and long-term learning transfer.

\textbf{Second, accessibility design could move beyond substituting vision with hearing and instead distribute information across low-attention modalities.} A recurring assumption in accessible interaction design is that when vision is unavailable, auditory channels can simply take its place—often resulting in designs that overload sound, implicitly treating hearing as a universal substitute for vision. Our findings challenge this: as M1 cautioned, prolonged reliance on auditory information can itself become harmful, \textit{“When you rely on hearing for a long time, your hearing will also be damaged.”} Rather than assuming "ears are the eyes for people with BLV," designers could distribute information across low-attention modalities—haptics, proprioceptive cues, subtle feedback—so no single channel bears the full cognitive and physiological load. For string learning, this means designing systems that support awareness without demanding constant listening, preserving musicians’ capacity to attend to sound as music rather than as interface output.

\textbf{Third, musicians with BLV’s expertise could be treated as a generative design resource rather than an edge case.} This work reinforces the value of designing \textit{with} and \textit{for} BLV expertise: participants’ reflections surfaced design ideas and tensions invisible in sighted-centered workflows. Treating musicians with BLV as knowledgeable practitioners, rather than edge cases, lets design research question foundational assumptions about perception and learning that remain obscured in vision-dominant pedagogical models.
Although grounded in upper-string learning, these implications extend to other accessibility contexts involving hands-busy, time-sensitive, and body-centered activity. In such settings, effective feedback may need to remain low-attention, practice-compatible, and integrated with users’ existing perceptual and motor routines, rather than introducing step-by-step cognitive overhead. In addition, effective design may depend less on translating visual information after the fact (e.g., turning a visual score into a Braille score, or a teacher’s gesture into a verbal description), and more on helping users perceive, interpret, and adjust ongoing bodily states without overloading a single sensory channel.

\section{Limitation and Future Direction}

This study has several limitations that also point toward important directions for future work. First, the population of violin and viola players with BLV is inherently small, particularly at advanced or professional levels of training—despite combining public recruitment with snowball sampling, our study involved a limited number of participants. This constraint reflects a structural reality rather than a recruitment shortcoming, and broad statistical generalizability was not a primary goal of this work.

Instead, we prioritized depth, diversity of perspective, and methodological triangulation. By combining practice videos, lesson observations, and interviews with both students and instructors, we captured rich, situated accounts of embodied learning that would be difficult to access through larger-scale or survey-based methods. The rarity of this population underscores the value of this material: participants’ accounts reflect years—often decades—of accumulated experience navigating visually grounded pedagogies without vision, valuable for understanding breakdowns in embodied learning that are otherwise taken for granted.

Second, our design phase engaged one musician with BLV (M2) for feedback. Scheduling constraints—M1 and M4's international location, M1's performance commitments, M4's health issues, and M3's atypical learning pathway—made broader participant iteration infeasible. However, design priorities remained grounded in instructor perspectives from Stage 1, and our author team included experienced string pedagogues who ensured concepts aligned with established teaching practice. Direct multi-participant design iteration remains important future work.

Future work will extend this research by developing interactive prototypes and engaging in iterative empirical evaluation with musicians with BLV and instructors, including short-term usability studies to assess perceptibility and cognitive load and longer-term deployments to examine how perceptual augmentation tools integrate into everyday practice routines and teacher–student relationships. In addition, future studies could explore how these design directions translate across instruments, skill levels, and cultural contexts, further examining how non-visual feedback systems might support embodied learning in other domains where vision is conventionally assumed.

\section{Conclusion}
This work examined how violin and viola players with BLV learn and practice within pedagogical systems that are deeply structured around vision, in addition to haptic and aural sensory perception. Through a qualitative, multi-method empirical study, we identified recurring difficulties in right-hand bow control, left-hand alignment and intonation, and score-related work during practice. These challenges reflect not only access barriers, but fundamental breakdowns in how continuous, embodied technique states are sensed and monitored when visual self-reference is unavailable. Building on these findings, we derived grounded design directions for how interactive systems might support non-visual, practice-compatible calibration. The resulting design concepts explore how otherwise invisible technique states—such as bow trajectory, grip tension, wrist–forearm alignment, and practice context—might be externalized through non-visual, low-attention feedback. Rather than enforcing fixed postures or replacing instruction, these designs could potentially support embodied calibration during independent practice while preserving flexibility and teacher judgment. These preliminary concepts require further development and validation with instructors and in practice contexts to assess whether they achieve these goals. Together, this work contributes empirically grounded insights into string learning with BLV. This work also proposes design directions that extend accessibility beyond information access toward perceptual augmentation and embodied learning support.

\section{Acknowledgments}
We would like to express our sincere gratitude to Qing Xiao of Carnegie Mellon University for valuable discussions that helped shape the initial ideas for this study. We are also grateful for his constructive feedback on the manuscript.

\bibliographystyle{ACM-Reference-Format}
\bibliography{bibili}

@article{lu2023there,
	title        = {“Why are there so many steps?”: Improving Access to Blind and Low Vision Music Learning through Personal Adaptations and Future Design Ideas},
	author       = {Lu, Leon and Cochrane, Karen Anne and Kang, Jin and Girouard, Audrey},
	year         = 2023,
	journal      = {ACM Transactions on Accessible Computing},
	publisher    = {ACM New York, NY},
	volume       = 16,
	number       = 3,
	pages        = {1--20}
}

@inproceedings{zhang2025accessibility,
	title        = {Accessibility and Social Inclusivity: A Literature Review of Music Technology for Blind and Low Vision People},
	author       = {Zhang, Shumeng and Masu, Raul and Bettega, Mela and Fan, Mingming},
	year         = 2025,
	booktitle    = {Proceedings of the 27th International ACM SIGACCESS Conference on Computers and Accessibility},
	pages        = {1--22}
}

@article{steriadis2003designing,
	title        = {Designing human-computer interfaces for quadriplegic people},
	author       = {Steriadis, Constantine E and Constantinou, Philip},
	year         = 2003,
	journal      = {ACM Transactions on Computer-Human Interaction (TOCHI)},
	publisher    = {ACM New York, NY, USA},
	volume       = 10,
	number       = 2,
	pages        = {87--118}
}

@article{howey1995technology,
	title        = {Technology for people with special needs: HCI design issues},
	author       = {Howey, Kate},
	year         = 1995,
	journal      = {CAMBRIDGE SERIES ON HUMAN COMPUTER INTERACTION},
	publisher    = {CAMBRIDGE UNIVERSITY PRESS},
	pages        = {343--360}
}

@phdthesis{moffatt2004designing,
	title        = {Designing technology for and with special populations: and exploration of participatory design with people with aphasia},
	author       = {Moffatt, Karyn},
	year         = 2004,
	school       = {university of british columbia}
}

@inproceedings{lu2023playing,
	title        = {Playing with feeling: Exploring vibrotactile feedback and aesthetic experiences for developing haptic wearables for blind and low vision music learning},
	author       = {Lu, Leon and Kang, Jin and Crispin, Chase and Girouard, Audrey},
	year         = 2023,
	booktitle    = {Proceedings of the 25th International ACM SIGACCESS Conference on Computers and Accessibility},
	pages        = {1--16}
}

@book{winograd1986understanding,
	title        = {Understanding computers and cognition: A new foundation for design},
	author       = {Winograd, Terry and Flores, Fernando and others},
	year         = 1986,
	publisher    = {Ablex publishing corporation Norwood, NJ},
	volume       = 335
}

@book{chen2006tactile,
	title        = {Tactile strategies for children who have visual impairments and multiple disabilities: Promoting communication and learning skills},
	author       = {Chen, Deborah and Downing, June E},
	year         = 2006,
	publisher    = {American Foundation for the Blind}
}

@article{pino2019teaching,
	title        = {Teaching--learning resources and supports in the music classroom: Key aspects for the inclusion of visually impaired students},
	author       = {Pino, Angela and Viladot, Laia},
	year         = 2019,
	journal      = {British Journal of Visual Impairment},
	publisher    = {SAGE Publications Sage UK: London, England},
	volume       = 37,
	number       = 1,
	pages        = {17--28}
}

@inproceedings{payne2025access,
	title        = {Access Beyond the Score: Understanding Notation Needs and Workflows of Low Vision Musicians},
	author       = {Payne, William Christopher and An, Yu Lee},
	year         = 2025,
	booktitle    = {Proceedings of the 27th International ACM SIGACCESS Conference on Computers and Accessibility},
	pages        = {1--14}
}

@article{baker2016perceptions,
	title        = {Perceptions of schooling, pedagogy and notation in the lives of visually-impaired musicians},
	author       = {Baker, David and Green, Lucy},
	year         = 2016,
	journal      = {Research Studies in Music Education},
	publisher    = {SAGE Publications Sage UK: London, England},
	volume       = 38,
	number       = 2,
	pages        = {193--219}
}

@inproceedings{ran2025users,
	title        = {How Users Who are Blind or Low Vision Play Mobile Games: Perceptions, Challenges, and Strategies},
	author       = {Ran, Zihe and Li, Xiyu and Xiao, Qing and Fan, Xianzhe and Li, Franklin Mingzhe and Wang, Yanyun and Lu, Zhicong},
	year         = 2025,
	booktitle    = {Proceedings of the 2025 CHI Conference on Human Factors in Computing Systems},
	pages        = {1--18}
}

@inproceedings{ran2025understanding,
	title        = {Understanding How Visually Impaired Players Socialize in Mobile Games},
	author       = {Ran, Zihe and Li, Xiyu and Xiao, Qing and Wang, Yanyun and Li, Franklin Mingzhe and Lu, Zhicong},
	year         = 2025,
	booktitle    = {Proceedings of the 27th International ACM SIGACCESS Conference on Computers and Accessibility},
	pages        = {1--16}
}

@article{palakshappa2006using,
	title        = {Using a multi-method qualitative approach to examine collaborative relationships},
	author       = {Palakshappa, Nitha and Ellen Gordon, Mary},
	year         = 2006,
	journal      = {Qualitative Market Research: An International Journal},
	publisher    = {Emerald Group Publishing Limited},
	volume       = 9,
	number       = 4,
	pages        = {389--403}
}

@article{allingham2021motor,
	title        = {Motor performance in violin bowing: Effects of attentional focus on acoustical, physiological and physical parameters of a sound-producing action},
	author       = {Allingham, Emma and Burger, Birgitta and W{\"o}llner, Clemens},
	year         = 2021,
	journal      = {Journal of New Music Research},
	publisher    = {Taylor \& Francis},
	volume       = 50,
	number       = 5,
	pages        = {428--446}
}

@article{rosenkranz2009regaining,
	title        = {Regaining motor control in musician's dystonia by restoring sensorimotor organization},
	author       = {Rosenkranz, Karin and Butler, Katherine and Williamon, Aaron and Rothwell, John C},
	year         = 2009,
	journal      = {Journal of Neuroscience},
	publisher    = {Society for Neuroscience},
	volume       = 29,
	number       = 46,
	pages        = {14627--14636}
}

@inproceedings{schoonderwaldt2011mastering,
	title        = {Mastering the violin: Motor learning in complex bowing skills},
	author       = {Schoonderwaldt, Erwin and Altenm{\"u}ller, Eckart},
	year         = 2011,
	booktitle    = {Proceedings of the International Symposium on Performance Science. Toronto, ON, Canada: AEC},
	pages        = {649--654}
}

@article{baader1999bimanual,
	title        = {Bimanual coordination in violin players: synchronization and preparation of finger movements},
	author       = {Baader, AP and Milani, P and Wiesendanger, M},
	year         = 1999,
	journal      = {Abstr. 28th Ann. Meet. Soc. Neurosci. Miami},
	volume       = 25
}

@incollection{wiesendanger2012fingering,
	title        = {Fingering and bowing in violinists: a motor control approach},
	author       = {Wiesendanger, Mario and Baader, Andreas and Kazennikov, Oleg},
	year         = 2012,
	booktitle    = {Music, motor control and the brain}
}

@book{galamian2013principles,
	title        = {Principles of violin playing and teaching},
	author       = {Galamian, Ivan and Thomas, Sally},
	year         = 2013,
	publisher    = {Courier Corporation}
}

@article{fischer1997basics,
	title        = {Basics: 300 exercises and practice routines for the violin},
	author       = {Fischer, Simon and others},
	year         = 1997,
	journal      = {(No Title)},
	publisher    = {Edition Peters}
}

@phdthesis{schoonderwaldt2009mechanics,
	title        = {Mechanics and acoustics of violin bowing: Freedom, constraints and control in performance},
	author       = {Schoonderwaldt, Erwin},
	year         = 2009,
	school       = {KTH}
}

@inproceedings{lu2022learning,
	title        = {Learning music blind: Understanding the application of technology to support BLV music learning},
	author       = {Lu, Leon},
	year         = 2022,
	booktitle    = {Proceedings of the 24th International ACM SIGACCESS Conference on Computers and Accessibility},
	pages        = {1--4}
}

@book{baker2017insights,
	title        = {Insights in Sound: Visually Impaired Musicians' Lives and Learning},
	author       = {Baker, David and Green, Lucy},
	year         = 2017,
	publisher    = {Routledge}
}

@inproceedings{langolff2000mfb,
	title        = {MFB (Music For the Blind): a software able to transcribe and create musical scores into Braille and to be used by blind persons},
	author       = {Langolff, Didier and Jessel, Nadine and Levy, Danny},
	year         = 2000,
	booktitle    = {6th ERCIM Workshop “User Interfaces for All” Short Paper},
	pages        = 6
}

@inproceedings{pedrini2020evaluating,
	title        = {Evaluating the accessibility of digital audio workstations for blind or visually impaired people},
	author       = {Pedrini, Gemma and Ludovico, Luca Andrea and Presti, Giorgio and others},
	year         = 2020,
	booktitle    = {Proceedings of the International Conference on Computer-Human Interaction Research and Applications (CHIRA 2020)},
	pages        = {225--232},
	organization = {Science and Technology Publications (SCITEPRESS)}
}

@inproceedings{payne2020blind,
	title        = {How blind and visually impaired composers, producers, and songwriters leverage and adapt music technology},
	author       = {Payne, William Christopher and Xu, Alex Yixuan and Ahmed, Fabiha and Ye, Lisa and Hurst, Amy},
	year         = 2020,
	booktitle    = {Proceedings of the 22nd International ACM SIGACCESS Conference on Computers and Accessibility},
	pages        = {1--12}
}

@inproceedings{lu2024we,
	title        = {" We Musicians Know How to Divide and Conquer": Exploring Multimodal Interactions To Improve Music Reading and Memorization for Blind and Low Vision Learners},
	author       = {Lu, Leon and Crispin, Chase and Girouard, Audrey},
	year         = 2024,
	booktitle    = {Proceedings of the 26th International ACM SIGACCESS Conference on Computers and Accessibility},
	pages        = {1--14}
}

@book{lazar2017research,
	title        = {Research methods in human-computer interaction},
	author       = {Lazar, Jonathan and Feng, Jinjuan Heidi and Hochheiser, Harry},
	year         = 2017,
	publisher    = {Morgan Kaufmann}
}

@incollection{ihas2023whole,
	title        = {Whole instrument approach: George Bornoff},
	author       = {Ihas, Dijana and Wilson, Miranda and McCormick, Gaelen},
	year         = 2023,
	booktitle    = {Teaching Violin, Viola, Cello, and Double Bass},
	publisher    = {Routledge},
	pages        = {165--168}
}

@article{mcgarry1984equal,
	title        = {Equal temperament, overtones, and the ear},
	author       = {McGarry, Robert J},
	year         = 1984,
	journal      = {Music educators journal},
	publisher    = {SAGE Publications Sage CA: Los Angeles, CA},
	volume       = 70,
	number       = 7,
	pages        = {54--56}
}

@book{duffin2007equal,
	title        = {How equal temperament ruined harmony (and why you should care)},
	author       = {Duffin, Ross W},
	year         = 2007,
	publisher    = {WW Norton \& Company}
}

@article{whitcomb2017intonation,
	title        = {Intonation on a string instrument: Three systems of tuning and temperament},
	author       = {Whitcomb, Benjamin},
	year         = 2017,
	journal      = {American String Teacher},
	publisher    = {SAGE Publications Sage CA: Los Angeles, CA},
	volume       = 67,
	number       = 2,
	pages        = {20--23}
}

@article{topper2002correcting,
	title        = {Correcting the right hand bow position for the student violinist and violist},
	author       = {Topper, Matson Alan},
	year         = 2002
}

@article{goldstein2000music,
	title        = {Music pedagogy for the blind},
	author       = {Goldstein, David},
	year         = 2000,
	journal      = {International Journal of Music Education},
	publisher    = {Sage Publications Sage UK: London, England},
	number       = 1,
	pages        = {35--39}
}

@article{chen2008pitch,
	title        = {Pitch and space maps of skilled cellists: accuracy, variability, and error correction},
	author       = {Chen, Jessie and Woollacott, Marjorie H and Pologe, Steven and Moore, George P},
	year         = 2008,
	journal      = {Experimental brain research},
	publisher    = {Springer},
	volume       = 188,
	number       = 4,
	pages        = {493--503}
}

@article{hafke2016violinists,
	title        = {Violinists' perceptions of and motor reactions to fundamental frequency shifts introduced in auditory feedback},
	author       = {Hafke-Dys, Honorata and Preis, Anna and Trojan, David},
	year         = 2016,
	journal      = {Acta Acustica united with Acustica},
	publisher    = {European Acoustics Association},
	volume       = 102,
	number       = 1,
	pages        = {155--158}
}

@article{zabanal2019effects,
	title        = {Effects of short-term practice with a tonic drone accompaniment on middle and high school violin and viola intonation},
	author       = {Zabanal, John-Rine A},
	year         = 2019,
	journal      = {String Research Journal},
	publisher    = {SAGE Publications Sage CA: Los Angeles, CA},
	volume       = 9,
	number       = 1,
	pages        = {51--61}
}

@article{morrison2011intonation,
	title        = {Intonation},
	author       = {Morrison, Steven and Fyk, Janina},
	year         = 2011,
	journal      = {The science \& psychology of music performance: Creative strategies for teaching and learning},
	publisher    = {Oxford University Press}
}

@article{mishra2000questions,
	title        = {Questions and answers: Research related to the teaching of string technique},
	author       = {Mishra, Jennifer},
	year         = 2000,
	journal      = {String Research Journal},
	publisher    = {SAGE Publications Sage CA: Los Angeles, CA},
	number       = 1,
	pages        = {9--35}
}

@article{shipps2014top,
	title        = {The top 10 greatest violin teachers and the top 10 violin influences in history},
	author       = {Shipps, Stephen},
	year         = 2014,
	journal      = {American String Teacher},
	publisher    = {SAGE Publications Sage CA: Los Angeles, CA},
	volume       = 64,
	number       = 4,
	pages        = {38--41}
}

@article{sarch1996violin,
	title        = {Violin Playing and Teaching in the'90s},
	author       = {Sarch, Kenneth L},
	year         = 1996,
	journal      = {American String Teacher},
	publisher    = {SAGE Publications Sage CA: Los Angeles, CA},
	volume       = 46,
	number       = 4,
	pages        = {52--54}
}

@article{berend1972interrelation,
	title        = {Interrelation of Weight and Motion in Violin Playing, Including the Art Of: Shifting},
	author       = {Berend, Margaret},
	year         = 1972,
	journal      = {American String Teacher},
	publisher    = {SAGE Publications Sage CA: Los Angeles, CA},
	volume       = 22,
	number       = 4,
	pages        = {22--25}
}

@article{visentin2015unraveling,
	title        = {Unraveling mysteries of personal performance style; biomechanics of left-hand position changes (shifting) in violin performance},
	author       = {Visentin, Peter and Li, Shiming and Tardif, Guillaume and Shan, Gongbing},
	year         = 2015,
	journal      = {PeerJ},
	publisher    = {PeerJ Inc.},
	volume       = 3,
	pages        = {e1299}
}

@article{mccrary2016effects,
	title        = {Effects of physical symptoms on muscle activity levels in skilled violinists},
	author       = {McCrary, J Matt and Halaki, Mark and Ackermann, Bronwen J},
	year         = 2016,
	journal      = {Medical problems of performing artists},
	publisher    = {Science \& Medicine, Inc.},
	volume       = 31,
	number       = 3,
	pages        = {125--131}
}

@book{rolland1974teaching,
	title        = {The teaching of action in string playing: Developmental and remedial techniques [for] violin and viola},
	author       = {Rolland, Paul and Mutschler, Marla},
	year         = 1974,
	publisher    = {Illinois String Research Associates},
	volume       = 1
}

@inproceedings{reel2004strings,
	title        = {Strings 101: Sound advice-fourteen steps toward improved intonation},
	author       = {Reel, J},
	year         = 2004,
	organization = {Strings}
}

@article{rolland1979movement,
	title        = {Movement in string playing: As it relates to the violin},
	author       = {Rolland, Paul},
	year         = 1979,
	journal      = {American String Teacher},
	publisher    = {SAGE Publications Sage CA: Los Angeles, CA},
	volume       = 29,
	number       = 1,
	pages        = {8--11}
}

@article{bergonzi1997effects,
	title        = {Effects of finger markers and harmonic context on performance of beginning string students},
	author       = {Bergonzi, Louis},
	year         = 1997,
	journal      = {Journal of Research in Music Education},
	publisher    = {SAGE Publications Sage CA: Los Angeles, CA},
	volume       = 45,
	number       = 2,
	pages        = {197--211}
}

@article{huovinen2024string,
	title        = {String teachers on the challenges of intonation: A report from Sweden},
	author       = {Huovinen, Erkki and Isabella Weng, Sheng-Ying},
	year         = 2024,
	journal      = {String Research Journal},
	publisher    = {Sage Publications Sage CA: Los Angeles, CA},
	pages        = 19484992241266528
}

@inproceedings{bowers2012logic,
	title        = {The logic of annotated portfolios: communicating the value of'research through design'},
	author       = {Bowers, John},
	year         = 2012,
	booktitle    = {Proceedings of the designing interactive systems conference},
	pages        = {68--77}
}

@inproceedings{zimmerman2007research,
	title        = {Research through design as a method for interaction design research in HCI},
	author       = {Zimmerman, John and Forlizzi, Jodi and Evenson, Shelley},
	year         = 2007,
	booktitle    = {Proceedings of the SIGCHI conference on Human factors in computing systems},
	pages        = {493--502}
}

@phdthesis{zhou2023right,
	title        = {The Right-Hand Technique in Violin Playing: a Pedagogical Study with a Focus on Coll{\'e} Action},
	author       = {Zhou, Yuhao},
	year         = 2023,
	school       = {University of Miami}
}

@book{auer1980violin,
	title        = {Violin playing as I teach it},
	author       = {Auer, Leopold},
	year         = 1980,
	publisher    = {Courier Corporation}
}

@incollection{ihas2023applications,
	title        = {Applications to modern teaching: A summary of concepts and pedagogical practices},
	author       = {Ihas, Dijana},
	year         = 2023,
	booktitle    = {Teaching Violin, Viola, Cello, and Double Bass},
	publisher    = {Routledge},
	pages        = {187--222}
}

@book{macleod2018teaching,
	title        = {Teaching strings in today's classroom: a guide for group instruction},
	author       = {MacLeod, Rebecca},
	year         = 2018,
	publisher    = {Routledge}
}

@article{hamann2009strategies,
	title        = {Strategies for teaching strings: Building a successful string and orchestra program},
	author       = {Hamann, Donald L and Gillespie, Robert},
	year         = 2009,
	journal      = {(No Title)}
}

@article{adams2008qualititative,
	title        = {A qualititative approach to HCI research},
	author       = {Adams, Anne and Lunt, Peter and Cairns, Paul},
	year         = 2008,
	publisher    = {Cambridge University Press}
}

@book{ellis1994selected,
	title        = {Selected band conductors' preparation to conduct selected band compositions},
	author       = {Ellis, Barry Len},
	year         = 1994,
	publisher    = {University of Illinois at Urbana-Champaign}
}

@book{hunsberger1992art,
	title        = {The art of conducting},
	author       = {Hunsberger, Donald and Ernst, Roy E},
	year         = 1992,
	publisher    = {McGraw-Hill}
}

@article{lane2006undergraduate,
	title        = {Undergraduate instrumental music education majors' approaches to score study in various musical contexts},
	author       = {Lane, Jeremy S},
	year         = 2006,
	journal      = {Journal of Research in Music Education},
	publisher    = {SAGE Publications Sage CA: Los Angeles, CA},
	volume       = 54,
	number       = 3,
	pages        = {215--230}
}

@article{terry2017thematic,
	title        = {Thematic analysis},
	author       = {Terry, Gareth and Hayfield, Nikki and Clarke, Victoria and Braun, Virginia and others},
	year         = 2017,
	journal      = {The SAGE handbook of qualitative research in psychology},
	publisher    = {SAGE Publications Ltd},
	volume       = 2,
	number       = {17-37},
	pages        = 25
}

@online{zweig_stringpedagogy_2025,
	title        = {Mimi Zweig StringPedagogy},
	author       = {Zweig, Mimi},
	year         = 2025,
	url          = {https://www.connollymusic.com/string-pedagogy},
	urldate      = {2026-01-18},
	organization = {Connolly Music Company}
}

@online{termeer_fretted_violin_nodate,
	title        = {The Fretted Violin: A Unique Instrument to Play!},
	author       = {Termeer, Julia},
	url          = {https://violinspiration.com/fretted-violin/},
	urldate      = {2026-01-18},
	website      = {Violinspiration}
}

@article{draper2022music,
	title        = {Music education for students with autism spectrum disorder in a full-inclusion context},
	author       = {Draper, Amanda R},
	year         = 2022,
	journal      = {Journal of Research in Music Education},
	publisher    = {SAGE Publications Sage CA: Los Angeles, CA},
	volume       = 70,
	number       = 2,
	pages        = {132--155}
}

@inproceedings{chen2026robots,
	title        = {Robots that evolve with us: Modular co-design for personalization, adaptability, and sustainability},
	author       = {Chen, Lingyun and Xiao, Qing and Zhang, Zitao and Blevis, Eli and {\v{S}}abanovi{\'c}, Selma},
	year         = 2026,
	booktitle    = {Proceedings of the 2026 CHI Conference on Human Factors in Computing Systems},
	pages        = {1--24}
}

@inproceedings{mankoff2010disability,
	title        = {Disability studies as a source of critical inquiry for the field of assistive technology},
	author       = {Mankoff, Jennifer and Hayes, Gillian R and Kasnitz, Devva},
	year         = 2010,
	booktitle    = {Proceedings of the 12th international ACM SIGACCESS conference on Computers and accessibility},
	pages        = {3--10}
}

@inproceedings{spiel2020nothing,
	title        = {Nothing about us without us: Investigating the role of critical disability studies in HCI},
	author       = {Spiel, Katta and Gerling, Kathrin and Bennett, Cynthia L and Brul{\'e}, Emeline and Williams, Rua M and Rode, Jennifer and Mankoff, Jennifer},
	year         = 2020,
	journal      = {ACM Transactions on Computer-Human Interaction (TOCHI)},
	booktitle    = {Extended Abstracts of the 2020 CHI Conference on Human Factors in Computing Systems},
	publisher    = {ACM},
	volume       = 26,
	number       = 6,
	pages        = {1--8}
}

@inproceedings{hofmann2020living,
	title        = {Living disability theory: Reflections on access, research, and design},
	author       = {Hofmann, Megan and Kasnitz, Devva and Mankoff, Jennifer and Bennett, Cynthia L},
	year         = 2020,
	booktitle    = {Proceedings of the 22nd International ACM SIGACCESS Conference on Computers and Accessibility},
	pages        = {1--13}
}

@article{schoonderwaldt2009violin,
	title        = {The Violinist's Sound Palette: Spectral Centroid, Pitch Flattening, and Anomalous Low Frequencies},
	author       = {Schoonderwaldt, Erwin},
	year         = 2009,
	journal      = {Acta Acustica united with Acustica},
	volume       = 95,
	number       = 5,
	pages        = {901--914}
}

@article{sigrist2013augmented,
	title        = {Augmented Visual, Auditory, Haptic, and Multimodal Feedback in Motor Learning: A Review},
	author       = {Sigrist, Roland and Rauter, Georg and Riener, Robert and Wolf, Peter},
	year         = 2013,
	journal      = {Psychonomic Bulletin \& Review},
	volume       = 20,
	pages        = {21--53}
}

@article{effenberg2005movement,
	title        = {Movement Sonification: Effects on Perception and Action},
	author       = {Effenberg, Alfred O.},
	year         = 2005,
	journal      = {IEEE MultiMedia},
	volume       = 12,
	number       = 2,
	pages        = {53--59}
}

@article{schaffert2011acoustic,
	title        = {An Investigation of Online Acoustic Information for Elite Rowers in On-Water Training Conditions},
	author       = {Schaffert, Nina and Mattes, Klaus and Effenberg, Alfred O.},
	year         = 2011,
	journal      = {Journal of Human Sport and Exercise},
	volume       = 6,
	number       = 2,
	pages        = {392--405}
}

@article{lieberman2007tikl,
	title        = {TIKL: Development of a Wearable Vibrotactile Feedback Suit for Improved Human Motor Learning},
	author       = {Lieberman, Jeff and Breazeal, Cynthia},
	year         = 2007,
	journal      = {IEEE Transactions on Robotics},
	volume       = 23,
	number       = 5,
	pages        = {919--926}
}

@inproceedings{spelmezan2012tactile,
	title        = {An Investigation into the Use of Tactile Instructions in Snowboarding},
	author       = {Spelmezan, Daniel},
	year         = 2012,
	booktitle    = {Proceedings of MobileHCI '12},
	pages        = {417--426}
}

@inproceedings{schonauer2012multimodal,
	title        = {Multimodal Motion Guidance: Techniques for Adaptive and Dynamic Feedback},
	author       = {Sch{\"o}nauer, Christian and Fukushi, Kenichiro and Olwal, Alex},
	year         = 2012,
	booktitle    = {Proceedings of ICMI '12},
	pages        = {133--140}
}

@article{vanderlinden2011musicjacket,
	title        = {MusicJacket: Combining Motion Capture and Vibrotactile Feedback to Teach Violin Bowing},
	author       = {van der Linden, Janet and Schoonderwaldt, Erwin and Bird, Jon and Johnson, Rose},
	year         = 2011,
	journal      = {IEEE Transactions on Instrumentation and Measurement},
	volume       = 60,
	number       = 1,
	pages        = {104--113}
}

@inproceedings{johnson2010musicjacket,
	title        = {MusicJacket: The Efficacy of Real-Time Vibrotactile Feedback for Learning to Play the Violin},
	author       = {Johnson, Rose and van der Linden, Janet and Rogers, Yvonne},
	year         = 2010,
	booktitle    = {CHI '10 Extended Abstracts on Human Factors in Computing Systems},
	pages        = {3475--3480}
}

@article{pardue2019realtime,
	title        = {Real-Time Aural and Visual Feedback for Improving Violin Intonation},
	author       = {Pardue, Laurel S. and McPherson, Andrew P.},
	year         = 2019,
	journal      = {Frontiers in Psychology},
	volume       = 10,
	pages        = 627
}

@article{dalmazzo2021realtime,
	title        = {Real-Time Sound and Motion Feedback for Violin Bow Technique Learning: A Controlled, Randomized Trial},
	author       = {Dalmazzo, David and Waddell, George and Ramirez, Rafael},
	year         = 2021,
	journal      = {Frontiers in Psychology},
	volume       = 12,
	pages        = 648479
}

@inproceedings{larkin2008sonification,
	title        = {Sonification of Bowing Features for String Instrument Training},
	author       = {Larkin, Oliver and Koerselman, Thijs and Ong, Bee and Ng, Kia},
	year         = 2008,
	booktitle    = {Proceedings of ICAD '08}
}

@article{giordano2021assessing,
	title        = {Assessing the Bowing Technique in Violin Beginners Using MIMU and Optical Proximity Sensors: A Feasibility Study},
	author       = {Giordano, Nicola and Dozio, Nicoletta and Frontoni, Emanuele and Roascio, Daniele and Zanoni, Massimiliano},
	year         = 2021,
	journal      = {Sensors},
	volume       = 21,
	number       = 17,
	pages        = 5817
}

@inproceedings{sharif2022should,
	title        = {Should I say “disabled people” or “people with disabilities”? Language preferences of disabled people between identity-and person-first language},
	author       = {Sharif, Ather and McCall, Aedan Liam and Bolante, Kianna Roces},
	year         = 2022,
	booktitle    = {Proceedings of the 24th international ACM SIGACCESS conference on computers and accessibility},
	pages        = {1--18}
}

@article{roulston2010reflective,
	title        = {Reflective interviewing: A guide to theory and practice},
	author       = {Roulston, Kathy},
	year         = 2010,
	publisher    = {Sage}
}

@book{patton2014qualitative,
	title        = {Qualitative research \& evaluation methods: Integrating theory and practice},
	author       = {Patton, Michael Quinn},
	year         = 2014,
	publisher    = {Sage publications}
}

@article{yin2026active,
	title        = {Active and passive social media use and self-stigmatisation among Chinese patients with gynecological disorders: the mediating role of social support},
	author       = {Yin, Menghan and Xiao, Qing},
	year         = 2026,
	journal      = {Behaviour \& Information Technology},
	publisher    = {Taylor \& Francis},
	pages        = {1--14}
}

@book{dunne2024speculative,
	title        = {Speculative Everything, With a new preface by the authors: Design, Fiction, and Social Dreaming},
	author       = {Dunne, Anthony and Raby, Fiona},
	year         = 2024,
	publisher    = {MIT press}
}

@inproceedings{chen20253r,
	title        = {3R (Robots, Rooms, Relationships): Speculative Homes, Sentient Machines, and the Future of Domesticity},
	author       = {Chen, Lingyun and Xiao, Qing and Siteri, Matyas Istvan and Blevis, Eli},
	year         = 2025,
	booktitle    = {Proceedings of the 2025 Conference on Creativity and Cognition},
	pages        = {211--223}
}

@book{emerson2011writing,
	title        = {Writing Ethnographic Fieldnotes},
	author       = {Emerson, Robert M. and Fretz, Rachel I. and Shaw, Linda L.},
	year         = 2011,
	publisher    = {University of Chicago Press},
	address      = {Chicago},
	edition      = 2
}

\end{document}